\documentclass[prd,reprint,onecolumn,notitlepage,preprintnumbers,superscriptaddress]{revtex4}
\usepackage{wrapfig}
\usepackage{amsfonts}
\usepackage{amsmath}
\usepackage{hyperref}
\usepackage{amssymb}
\usepackage[english]{babel}
\usepackage{graphicx}
\usepackage{epsfig}
\usepackage{bm}
\usepackage{longtable}
\usepackage{verbatim}
\usepackage{longtable}
\usepackage{color}
\usepackage{subfigure}
\usepackage[utf8]{inputenc}

\newcommand{\mathsym}[1]{{}}

\input{tcilatex}
\begin{document}

\title{Fermion masses and mixings in the 3-3-1 model with right-handed
neutrinos based on the $S_3$ flavor symmetry.}
\author{A. E. C\'arcamo Hern\'andez${}^{a}$}
\email{antonio.carcamo@usm.cl}
\author{R. Martinez${}^{{b}}$}
\email{remartinezm@unal.edu.co}
\author{F. Ochoa${}^{{b}}$}
\email{faochoap@unal.edu.co}
\affiliation{$^{{a}}$Universidad T\'{e}cnica Federico Santa Mar\'{\i}a, Casilla 110-V,
Valpara\'{\i}so, Chile,\\
$^{{b}}$Universidad Nacional de Colombia, Departamento de F\'{\i}sica,
Ciudad Universitaria, Bogot\'{a} D.C., Colombia. }
\date{\today }

\begin{abstract}
We propose a 3-3-1 model where the $SU(3)_{C}\otimes SU(3)_{L}\otimes
U(1)_{X}$ symmetry is extended by $S_{3}\otimes Z_{3}\otimes Z_{3}^{\prime
}\otimes Z_{8}\otimes Z_{16}$ and the scalar spectrum is enlarged by extra $%
SU(3)_{L}$ singlet scalar fields. The model successfully describes the
observed SM fermion mass and mixing pattern. In this framework, the light
active neutrino masses arise via an inverse seesaw mechanism and the
observed charged fermion mass and quark mixing hierarchy is a consequence of
the $Z_{3}\otimes Z_{3}^{\prime }\otimes Z_{8}\otimes Z_{16}$ symmetry
breaking at very high energy. The obtained physical observables for both
quark and lepton sectors are compatible with their experimental values. The
model predicts the effective Majorana neutrino mass parameter of
neutrinoless double beta decay to be $m_{\beta \beta }=$ 4 and 48 meV for
the normal and the inverted neutrino spectra, respectively. Furthermore, we
found a leptonic Dirac CP violating phase close to $\frac{\pi }{2}$ and a
Jarlskog invariant close to about $3\times 10^{-2}$ for both normal and
inverted neutrino mass hierarchy.
\end{abstract}

\maketitle



\section{Introduction}

After the discovery of the $126$ GeV Higgs boson by ATLAS and CMS
collaborations at CERN Large Hadron Collider (LHC) \cite%
{Aad:2012tfa,Chatrchyan:2012xdj}, the vacancy of the Higgs boson needed for
the completion of the Standard Model (SM) at the Fermi scale has been filled
and the weak gauge bosons mass generation mechanism has also been confirmed.
Despite LHC experiments indicate that the decay modes of the new scalar
state are SM like, there is still room for new extra scalar states, whose
search are an essential task of the LHC experiments. Furthermore, despite
the great consistency of the SM predictions with the experimental data,
there are several aspects that the SM do not explain, some of them are the
observed hierarchy among charged fermion masses and quark mixing angles, the
tiny neutrino masses and the smallness of the quark mixing angles, which
contrast with the sizeable leptonic mixing ones. The global fits of the
available data from the Daya Bay \cite{An:2012eh}, T2K \cite{Abe:2011sj},
MINOS \cite{Adamson:2011qu}, Double CHOOZ \cite{Abe:2011fz} and RENO \cite%
{Ahn:2012nd} neutrino oscillation experiments, constrain the neutrino mass
squared splittings and mixing parameters \cite{Forero:2014bxa}. It is a well
established experimental fact that the observed hierarchy of charged fermion
masses goes over a range of five orders of magnitude in the quark sector and
that there are six orders of magnitude between the neutrino mass scale and
the electron mass. Accommodating the charged fermion masses in the SM
requires an unnatural tunning among its different Yukawa couplings.
Furthermore, experiments with solar, atmospheric and reactor neutrinos \cite%
{Agashe:2014kda,An:2012eh,Abe:2011sj,Adamson:2011qu,Abe:2011fz,Ahn:2012nd}
have brought evidence of neutrino oscillations caused by nonzero mass. All
these unexplained issues strongly indicate that new physics have to be
invoked to address the fermion puzzle of the SM. 

\begin{table}[tbh]
\begin{tabular}{|c|c|c|c|c|c|}
\hline
Parameter & $\Delta m_{21}^{2}$($10^{-5}$eV$^2$) & $\Delta m_{31}^{2}$($%
10^{-3}$eV$^2$) & $\left( \sin ^{2}\theta _{12}\right) _{\exp }$ & $\left(
\sin ^{2}\theta _{23}\right) _{\exp }$ & $\left( \sin ^{2}\theta
_{13}\right) _{\exp }$ \\ \hline
Best fit & $7.60$ & $2.48$ & $0.323$ & $0.567$ & $0.0234$ \\ \hline
$1\sigma $ range & $7.42-7.79$ & $2.41-2.53$ & $0.307-0.339$ & $0.439-0.599$
& $0.0214-0.0254$ \\ \hline
$2\sigma $ range & $7.26-7.99$ & $2.35-2.59$ & $0.292-0.357$ & $0.413-0.623$
& $0.0195-0.0274$ \\ \hline
$3\sigma $ range & $7.11-8.11$ & $2.30-2.65$ & $0.278-0.375$ & $0.392-0.643$
& $0.0183-0.0297$ \\ \hline
\end{tabular}%
\caption{Range for experimental values of neutrino mass squared splittings
and leptonic mixing parameters, taken from Ref. \protect\cite{Forero:2014bxa}%
, for the case of normal hierarchy.}
\end{table}
\begin{table}[tbh]
\begin{tabular}{|c|c|c|c|c|c|}
\hline
Parameter & $\Delta m_{21}^{2}$($10^{-5}$eV$^{2}$) & $\Delta m_{13}^{2}$($%
10^{-3}$eV$^{2}$) & $\left( \sin ^{2}\theta _{12}\right) _{\exp }$ & $\left(
\sin ^{2}\theta _{23}\right) _{\exp }$ & $\left( \sin ^{2}\theta
_{13}\right) _{\exp }$ \\ \hline
Best fit & $7.60$ & $2.38$ & $0.323$ & $0.573$ & $0.0240$ \\ \hline
$1\sigma $ range & $7.42-7.79$ & $2.32-2.43$ & $0.307-0.339$ & $0.530-0.598$
& $0.0221-0.0259$ \\ \hline
$2\sigma $ range & $7.26-7.99$ & $2.26-2.48$ & $0.292-0.357$ & $0.432-0.621$
& $0.0202-0.0278$ \\ \hline
$3\sigma $ range & $7.11-8.11$ & $2.20-2.54$ & $0.278-0.375$ & $0.403-0.640$
& $0.0183-0.0297$ \\ \hline
\end{tabular}%
\caption{Range for experimental values of neutrino mass squared splittings
and leptonic mixing parameters, taken from Ref. \protect\cite{Forero:2014bxa}%
, for the case of inverted hierarchy.}
\label{IH}
\end{table}

\quad The aforementioned flavour puzzle, not understood in the context of
the SM, motivates extensions of the Standard Model that explain the fermion
mass and mixing patterns. From the phenomenological point of view, it is
possible to describe some features of the mass hierarchy by assuming Yukawa
matrices with texture zeroes \cite%
{Fritzsch:1977za,Fukuyama:1997ky,Du:1992iy,Barbieri:1994kw,Peccei:1995fg,Fritzsch:1999ee,Roberts:2001zy,Nishiura:2002ei,deMedeirosVarzielas:2005ax,Carcamo:2006dp,Kajiyama:2007gx,CarcamoHernandez:2010im,Branco:2010tx,Leser:2011fz,Gupta:2012dma,CarcamoHernandez:2012xy,Hernandez:2013mcf,Pas:2014bra,Hernandez:2014hka,Hernandez:2014zsa,Nishiura:2014psa,Frank:2014aca,Ghosal:2015lwa,Sinha:2015ooa,Nishiura:2015qia,Samanta:2015oqa,Gautam:2015kya,Pas:2015hca,Hernandez:2015hrt}%
. A very promising approach is the use of discrete flavor groups, which have
been considered in several models to explain the fermion masses and mixing
(see Refs. \cite{Ishimori:2010au,Altarelli:2010gt,King:2013eh,King:2014nza}
for recent reviews on flavor symmetries). Models with spontaneously broken
flavor symmetries may also produce hierarchical mass structures. Recently,
discrete groups such as $A_{4}$ \cite%
{Ma:2001dn,He:2006dk,Chen:2009um,Ahn:2012tv,Memenga:2013vc,Felipe:2013vwa,Varzielas:2012ai,Ishimori:2012fg,King:2013hj,Hernandez:2013dta,Babu:2002dz,Altarelli:2005yx,Morisi:2013eca,Altarelli:2005yp,Kadosh:2010rm,Kadosh:2013nra,delAguila:2010vg,Campos:2014lla,Vien:2014pta,Hernandez:2015tna}%
, $S_{3}$ \cite%
{Kubo:2003pd,Kobayashi:2003fh,Chen:2004rr,Mondragon:2007af,Mondragon:2008gm,Bhattacharyya:2010hp,Dong:2011vb,Dias:2012bh,Meloni:2012ci,Canales:2012dr,Canales:2013cga,Ma:2013zca,Kajiyama:2013sza,Hernandez:2013hea,Ma:2014qra,Hernandez:2014vta,Hernandez:2014lpa,Gupta:2014nba,Hernandez:2015dga,Hernandez:2015zeh,Hernandez:2016rbi}%
, $S_{4}$ \cite%
{Mohapatra:2012tb,BhupalDev:2012nm,Varzielas:2012pa,Ding:2013hpa,Ishimori:2010fs,Ding:2013eca,Hagedorn:2011un,Campos:2014zaa,Dong:2010zu,VanVien:2015xha,Arbelaez:2016mhg}%
, $D_4$ \cite%
{Frampton:1994rk,Grimus:2003kq,Grimus:2004rj,Frigerio:2004jg,Babu:2004tn,Adulpravitchai:2008yp,Ishimori:2008gp,Hagedorn:2010mq,Meloni:2011cc,Vien:2013zra}%
, $Q_6$ \cite%
{Kawashima:2009jv,Kaburaki:2010xc,Babu:2011mv,Gomez-Izquierdo:2013uaa}, $T_7$
\cite%
{Luhn:2007sy,Hagedorn:2008bc,Cao:2010mp,Luhn:2012bc,Kajiyama:2013lja,Bonilla:2014xla,Vien:2014gza,Vien:2015koa,Hernandez:2015cra,Arbelaez:2015toa}%
, $T_{13}$ \cite{Ding:2011qt,Hartmann:2011dn,Hartmann:2011pq,Kajiyama:2010sb}%
, $T^{\prime }$ \cite%
{Aranda:2000tm,Aranda:2007dp,Chen:2007afa,Frampton:2008bz,Eby:2011ph,Frampton:2013lva,Chen:2013wba}%
, $\Delta(27)$ \cite%
{Ma:2007wu,Varzielas:2012nn,Bhattacharyya:2012pi,Ma:2013xqa,Nishi:2013jqa,Varzielas:2013sla,Aranda:2013gga,Ma:2014eka,Abbas:2014ewa,Abbas:2015zna,Varzielas:2015aua,Bjorkeroth:2015uou,Chen:2015jta,Vien:2016tmh,Hernandez:2016eod}
and $A_5$ \cite%
{Everett:2008et,Feruglio:2011qq,Cooper:2012bd,Varzielas:2013hga,Gehrlein:2014wda,Gehrlein:2015dxa,DiIura:2015kfa,Ballett:2015wia,Gehrlein:2015dza,Turner:2015uta,Li:2015jxa}
have been considered to explain the observed pattern of fermion masses and
mixings. In particular the $S_{3}$ flavor symmetry is a very good candidate
for explaining the prevailing pattern of fermion masses and mixing. The $%
S_{3}$ discrete symmetry is the smallest non-Abelian discrete symmetry group
having three irreducible representations (irreps), explicitly two singlets
and one doublet irreps. The $S_3$ discrete symmetry was used as a flavor
symmetry for the first time in Ref. \cite{Pakvasa:1977in}. The different
models based on discrete flavor symmetries, have as a common issue the
breaking of the flavour symmetry so that the observed data be naturally
produced. The breaking of the flavour symmetry takes place when the scalar
fields acquire vacuum expectation values.

\quad Besides that, another of the greatest misteries in particle physics is
the existence of three fermion families at low energies. The origin of the
family structure of the fermions can be addressed in family dependent models
where a symmetry distinguish fermions of different families. One explanation
to this issue can be provided by the models based on the gauge symmetry $%
SU(3)_{c}\otimes SU(3)_{L}\otimes U(1)_{X}$, also called 3-3-1 models, which
introduce a family non-universal $U(1)_{X}$ symmetry \cite%
{Georgi:1978bv,Valle:1983dk,Pisano:1991ee,Montero:1992jk,Foot:1992rh,Frampton:1992wt,Ng:1992st,Duong:1993zn,Hoang:1996gi,Hoang:1995vq,Foot:1994ym,Martinez:2001mu,Sanchez:2001ua,Diaz:2003dk,Diaz:2004fs,Dias:2004dc,Dias:2005yh,Dias:2005jm,Ochoa:2005ch,CarcamoHernandez:2005ka,Salazar:2007ym,Benavides:2009cn,Dias:2010vt,Dias:2012xp,Alvarado:2012xi,Catano:2012kw,Hernandez:2013mcf,Hernandez:2014lpa,Vien:2014pta,Hernandez:2014vta,Boucenna:2014ela,Boucenna:2014dia,Vien:2014gza,Phong:2014ofa,Boucenna:2015zwa,Hernandez:2015cra,DeConto:2015eia,Correia:2015tra,Dong:2015rka,Hernandez:2015tna,Okada:2015bxa,Binh:2015cba,Hue:2015fbb,Benavides:2015afa,Boucenna:2015pav,Hernandez:2015ywg,Dong:2015dxw,Cao:2015scs,Martinez:2016ztt,Borges:2016nne,Okada:2016whh,Fonseca:2016xsy,Fonseca:2016tbn}%
. These models have a number of phenomenological advantages. First of all,
the three family structure in the fermion sector can be understood in the
3-3-1 models from the cancellation of chiral anomalies and asymptotic
freedom in QCD. Secondly, the fact that the third family is treated under a
different representation, can explain the large mass difference between the
heaviest quark family and the two lighter ones. Third, these models contain
a natural Peccei-Quinn symmetry, necessary to solve the strong-CP problem 
\cite{Pal:1994ba,Dias:2002gg,Dias:2003zt,Dias:2003iq}. Finally, 3-3-1 models
including heavy sterile neutrinos have cold dark matter candidates as weakly
interacting massive particles (WIMPs) \cite%
{Mizukoshi:2010ky,Dias:2010vt,Alvares:2012qv,Cogollo:2014jia}. Besides that,
the 3-3-1 models can explain the $2$ TeV diboson excess found by ATLAS \cite%
{Cao:2015lia}. When the electric charge in the 3-3-1 models is defined in
the linear combination of the $SU(3)_{L}$ generators $T_{3}$ and $T_{8}$, it
is a free parameter, independent of the anomalies ($\beta $). The choice of
this parameter defines the charge of the exotic particles. Choosing $\beta =-%
\frac{1}{\sqrt{3}}$, the third component of the weak lepton triplet is a
neutral field $\nu _{R}^{C}$ which allows to build the Dirac matrix with the
usual field $\nu _{L}$ of the weak doublet. If one introduces a sterile
neutrino $N_{R}$ in the model, then it is possible to generate light
neutrino masses via inverse seesaw mechanism. The 3-3-1 models with $\beta =-%
\frac{1}{\sqrt{3}}$ have the advantange of providing an alternative
framework to generate neutrino masses, where the neutrino spectrum includes
the light active sub-eV scale neutrinos as well as sterile neutrinos which
could be dark matter candidates if they are light enough or candidates for
detection at the LHC, if their masses are at the TeV scale. This interesting
feature make the 3-3-1 models very interesting since if the TeV scale
sterile neutrinos are found at the LHC, these models can be very strong
candidates for unraveling the mechanism responsible for electroweak symmetry
breaking.

\quad In the 3-3-1 models, one heavy triplet field with a Vacuum Expectation
Value (VEV) at high energy scale $\nu _{\chi }$, breaks the symmetry $%
SU(3)_{L}\otimes U(1)_{X}$ into the SM electroweak group $SU(2)_{L}\otimes
U(1)_{Y}$, while the another two lighter triplets with VEVs at the
electroweak scale $\upsilon _{\rho }$ and $\upsilon _{\eta }$, trigger the
Electroweak Symmetry Breaking \cite{Hernandez:2013mcf}. 
Besides that, the 3-3-1 model could possibly explain the excess of events in
the $h\rightarrow \gamma \gamma $ decay, recently observed at the LHC, since
the heavy exotic quarks, the charged Higges and the heavy charged gauge
bosons contribute to this process. On the other hand, the 3-3-1 model
reproduces an specialized Two Higgs Doublet Model type III (2HDM-III) in the
low energy limit, where both electroweak triplets $\rho $ and $\eta $ are
decomposed into two hypercharge-one $SU(2)_{L}$ doublets plus charged and
neutral singlets. Thus, like the 2HDM-III, the 3-3-1 model can predict huge
flavor changing neutral currents (FCNC) and CP-violating effects, which are
severely suppressed by experimental data at electroweak scales. In the
2HDM-III, for each quark type, up or down, there are two Yukawa couplings.
One of the Yukawa couplings is for generating the quark masses, and the
other one produces the flavor changing couplings at tree level. One way to
remove both the huge FCNC and CP-violating effects, is by imposing discrete
symmetries, obtaining two types of 3-3-1 models (type I and II models),
which exhibit the same Yukawa interactions as the 2HDM type I and II at low
energy where each fermion is coupled at most to one Higgs doublet. In the
3-3-1 model type I, one Higgs electroweak triplet (for example, $\rho $)
provide masses to the phenomenological up- and down-type quarks,
simultaneously. In the type II, one Higgs triplet ($\eta $) gives masses to
the up-type quarks and the other triplet ($\rho $) to the down-type quarks 
\cite{Hernandez:2013mcf}.

\quad It is noteworthy the $S_3$ flavor symmetry was implemented for the
first time in the 3-3-1 model of Ref. \cite{Dong:2011vb}. That model
introduces a new $\mathrm{U}(1)_{\mathcal{L}}$ lepton global symmetry,
responsible for lepton number and lepton parity. That lepton parity symmetry
suppresses the mixing between ordinary quarks and exotic quarks.
Furthermore, the $\mathrm{U}(1)_{\mathcal{L}}$ new lepton global symmetry
enforces to have different scalar fields in the Yukawa interactions for
charged lepton, neutrino and quark sectors. The scalar sector of that model
includes six $SU(3)_L$ scalar triplets and four $SU(3)_L$ scalar
antisextets. The $\mathrm{SU}(3)_{C}\otimes \mathrm{SU}(3)_{L}\otimes 
\mathrm{U}(1)_{X}\otimes\mathrm{U}(1)_{\mathcal{L}}\otimes S_3$ assignments
of the fermion sector of the the aforementioned model, require that these 6 $%
SU(3)_L$ scalar triplets be distributed as follows, 3 for the quark sector,
2 for the charged lepton sector and 1 for the neutrino sector. Furthermore
the 4 $SU(3)_L$ scalar antisextets are needed to implement a type II seesaw
mechanism. In that model, light active neutrino masses are generated from
type-I and type-II seesaw mechanisms, mediated by three heavy right handed
Majorana neutrinos and four $SU(3)_{L}$ scalar antisextets, respectively.
Since the Yukawa terms of that model are renormalizable, to explain the SM
charged fermion mass pattern, one needs to impose a strong hierarchy among
the charged fermion Yukawa couplings of the model. Furthermore, the work
described in Ref. \cite{Dong:2011vb} is mainly focused on the lepton sector,
while in the quark sector, the obtained quark mass matrices are diagonal and
the quark mixing matrix is trivial. 

\quad Recently two of us proposed a $SU(3)_{C}\times SU(3)_{L}\times
U(1)_{X}\otimes S_{3}\otimes Z_{2}\otimes Z_{4}\otimes Z_{12}$ model \cite%
{Hernandez:2014vta}, with a scalar sector composed of three $SU(3)_{L}$
scalar triplets and seven $SU(3)_{L}$ scalar singlets, that successfully
accounts for quark masses and mixings. In that model, all observables in the
quark sector are in excellent agreement with the experimental data,
excepting $\bigl|V_{td}\bigr|$, which turns out to be larger by a factor $%
\sim 1.3$ than its corresponding experimental value, and naively deviated 8
sigma away from it. That model has the following drawbacks: $\bigl|V_{td}%
\bigr|$ is deviated 8 sigma away from its experimental value, a $S_{3}$ soft
breaking term has to be introduced by hand in the low energy scalar
potential in order to fullfill its minimization equations, the top quark
mass arises from a five dimensional Yukawa term and lepton masses and
mixings are not addressed.

\quad It is interesting to find an alternative and better explanation for
the SM fermion mass and mixing hierarchy than the ones considered in Refs. 
\cite{Dong:2011vb,Hernandez:2014vta}. To this end we 
propose a multiscalar singlet extension of the $SU(3)_{C}\times
SU(3)_{L}\times U(1)_{X}$ model with right handed neutrinos, where $\beta =-%
\frac{1}{\sqrt{3}}$ and an extra 
\mbox{$S_{3}\otimes Z_{3}\otimes Z_{3}^{\prime}\otimes
Z_{8}\otimes Z_{16}$} discrete group, extends the symmetry of the model and
fifteen very heavy $SU(3)_{L}$ singlet scalar fields are added with the aim
to generate viable textures for the fermion sector, that successfully
describe the observed SM fermion mass and mixing pattern. Let us note that
whereas the scalar sector of our model only has three $SU(3)_{L}$ scalar
triplets and fifteen $SU(3)_{L}$ scalar singlets, the scalar sector of the $%
S_{3}$ flavour 3-3-1 model of Ref. \cite{Dong:2011vb} has six $SU(3)_{L}$
scalar triplets and four $SU(3)_{L}$ scalar antisextets. Whereas in the
model of Ref. \cite{Dong:2011vb}, the quark mixing matrix is equal to the
identity, in our model the quark mixing matrix is in excellent agreement
with the low energy quark flavor data. In our model, the obtained physical
observables in the quark and lepton sector are consistent with the
experimental data. Our model at low energies reduces to the 3-3-1 model with
right handed neutrinos, where $\beta =-\frac{1}{\sqrt{3}}$. Furthermore, our
current model does not include the $\mathrm{U}(1)_{\mathcal{L}}$ new lepton
global symmetry presented in the $S_{3}$ flavor 3-3-1 model of Ref. \cite%
{Dong:2011vb}. Unlike the $S_{3}$ flavor 3-3-1 model of Ref. \cite%
{Dong:2011vb}, in our current 3-3-1 model, the charged fermion mass and
quark mixing pattern can successfully be accounted for, by having all Yukawa
couplings of order unity and arises from the breaking of the $Z_{3}\otimes
Z_{3}^{\prime }\otimes Z_{8}\otimes Z_{16}$ discrete group at very high
energy, triggered by $SU(3)_{L}$ scalar singlets acquiring vacuum
expectation values much larger than the TeV scale. Despite our current model
has more $SU(3)_{L}$ scalar singlets than the model that two of us have
recently proposed in Ref. \cite{Hernandez:2014vta}, our current model
addresses both the quark and lepton sectors and does not have the
aforementioned drawbacks of the model of Ref. \cite{Hernandez:2014vta}.
Because of the aforementioned reasons, our current model represents an
important improvement over the previously studied scenarios \cite%
{Dong:2011vb,Hernandez:2014vta}. 
The particular role of each additional scalar field and the corresponding
particle assignments under the symmetry group of the model under
consideration are explained in details in Sec. \ref{331model}. The model we
are building with the aforementioned discrete symmetries, preserves the
content of particles of the 3-3-1 model with $\beta =-\frac{1}{\sqrt{3}}$,
but we add fifteen additional very heavy $SU(3)_{L}$ singlet scalar fields,
with quantum numbers that allow to build Yukawa terms invariant under the
local and discrete groups. This generates the right textures that
successfully account for SM fermion masses and mixings. We assume that the
Majorana neutrinos have very small masses, 
implying that the small active neutrino masses are generated via an inverse
seesaw mechanism. This mechanism for the generation of the light active
neutrino masses differs from the one implemented in the $S_{3}$ flavor 3-3-1
model of Ref. \cite{Dong:2011vb}, where the light active neutrinos get their
masses from type I and type II seesaw mechanisms. 

\quad The paper is outlined as follows. In Sec. \ref{331model} we explain
some theoretical aspects of the 3-3-1 model with $\beta =-\frac{1}{\sqrt{3}}$
and its particle content, as well as the particle assignments under doublet
and singlet $S_{3}$ representations, in particular in the fermionic and
scalar sector. 
The low energy scalar potential of our model is discussed in Sec \ref%
{scalarpotential}. In Sec. \ref{leptonmassesandmixing} we focus on the
discussion of lepton masses and mixing and give our corresponding results.
In Sec. \ref{quarkmassesandmixing}, we present our results in terms of quark
masses and mixing, which is followed by a numerical analysis. Conclusions
are given Sec. \ref{conclusions}. In the appendices we present several
technical details: Appendix \ref{S3} gives a brief description of the $S_{3}$
group; Appendix \ref{stability} shows a discussion of the stability
conditions of the low energy scalar potential.

\section{The model}

\label{331model}

\subsection{Particle content}

The first 3-3-1 model with right handed Majorana neutrinos in the $SU(3)_{L}$
lepton triplet was considered in \cite{Montero:1992jk}. However that model
cannot describe the observed pattern of SM fermion masses and mixings, due
to the unexplained hierarchy among the large number of Yukawa couplings in
the model. Below we consider a $SU(3)_{C}\otimes SU(3)_{L}\otimes
U(1)_{X}\otimes S_{3}\otimes Z_{3}\otimes Z_{3}^{\prime }\otimes
Z_{8}\otimes Z_{16}$ multiscalar singlet extension of the 3-3-1 model with
right-handed neutrinos, which successfully describes the SM fermion mass and
mixing pattern. In our model the full symmetry $\mathcal{G}$ 
is spontaneously broken in three steps as follows: 
\begin{eqnarray}
&&\mathcal{G}=SU(3)_{C}\otimes SU\left( 3\right) _{L}\otimes U\left(
1\right) _{X}\otimes S_{3}\otimes Z_{3}\otimes Z_{3}^{\prime }\otimes
Z_{8}\otimes Z_{16}{\xrightarrow{\Lambda _{int}}}  \notag \\
&&\hspace{7mm}SU(3)_{C}\otimes SU\left( 3\right) _{L}\otimes U\left(
1\right) _{X}\otimes Z_{3}{\xrightarrow{v_{\chi }}}SU(3)_{C}\otimes SU\left(
2\right) _{L}\otimes U\left( 1\right) _{Y}{\xrightarrow{v_{\eta },v_{\rho }}}
\notag \\
&&\hspace{7mm}SU(3)_{C}\otimes U\left( 1\right) _{Q},  \label{Group}
\end{eqnarray}%
where the hierarchy $v_{\eta },v_{\rho }\ll v_{\chi }\ll \Lambda _{int}$
among the symmetry breaking scales is fullfilled.


\quad The electric charge in our 3-3-1 model is defined as \cite{Dong:2010zu}%
: 
\begin{equation}
Q=T_{3}-\frac{1}{\sqrt{3}}T_{8}+XI,
\end{equation}%
where $T_3$ and $T_8$ are the $SU(3)_L$ diagonal generators, $I$ is the $%
3\times 3$ identity matrix and $X$ the $U(1)_X$ charge. 

\quad Two families of quarks are grouped in a $3^{\ast }$ irreducible
representations (irreps), as required from the $SU(3)_{L}$ anomaly
cancellation. Furthermore, from the quark colors, it follows that the number
of $3^{\ast }$ irreducible representations is six. The other family of
quarks is grouped in a $3$ irreducible representation. Moreover, there are
six $3$ irreps taking into account the three families of leptons.
Consequently, the $SU(3)_{L}$ representations are vector like and do not
contain anomalies. The quantum numbers for the fermion families are assigned
in such a way that the combination of the $U(1)_{X}$ representations with
other gauge sectors is anomaly free. Therefore, the anomaly cancellation
requirement implies that quarks are unified in the following $%
(SU(3)_{C},SU(3)_{L},U(1)_{X})$ left- and right-handed representations: 
\begin{align}
Q_{L}^{1,2}& =%
\begin{pmatrix}
D^{1,2} \\ 
-U^{1,2} \\ 
J^{1,2} \\ 
\end{pmatrix}%
_{L}:(3,3^{\ast },0),\hspace{1cm}Q_{L}^{3}=%
\begin{pmatrix}
U^{3} \\ 
D^{3} \\ 
T \\ 
\end{pmatrix}%
_{L}:(3,3,1/3),  \label{fermion_spectrumleft} \\
& 
\begin{array}{c}
D_{R}^{1,2,3}:(3,1,-1/3), \\ 
J_{R}^{1,2}:(3,1,-1/3),%
\end{array}%
\hspace{0.7cm}%
\begin{array}{c}
U_{R}^{1,2,3}:(3,1,2/3), \\ 
T_{R}:(3,1,2/3).%
\end{array}
\label{fermion_spectrumright}
\end{align}%
Here $U_{L}^{i}$ and $D_{L}^{i}$ ($i=1,2,3$) are the left handed up- and
down-type quarks in the flavor basis. The right handed SM\ quarks $U_{R}^{i}$
and $D_{R}^{i}$ ($i=1,2,3$) and right handed exotic quarks $T_{R}$ and $%
J_{R}^{1,2}$ are assigned into $SU(3)_{L}$ singlets representations, so that
their $U(1)_{X}$ quantum numbers correspond to their electric charges.

Furthermore, cancellation of anomalies implies that leptons are grouped in
the following $(SU(3)_{C},SU(3)_{L},U(1)_{X})$ left- and right-handed
representations: 
\begin{align}
L_{L}^{1,2,3}& =%
\begin{pmatrix}
\nu ^{1,2,3} \\ 
e^{1,2,3} \\ 
(\nu ^{1,2,3})^{c} \\ 
\end{pmatrix}%
_{L}:(1,3,-1/3), \\
& 
\begin{array}{c}
e_{R}:(1,1,-1), \\ 
N_{R}^{1}:(1,1,0), \\ 
\end{array}%
\hspace{0.7cm}%
\begin{array}{c}
\mu _{R}:(1,1,-1), \\ 
N_{R}^{2}:(1,1,0), \\ 
\end{array}%
\hspace{0.7cm}%
\begin{array}{c}
\tau _{R}:(1,1,-1), \\ 
N_{R}^{3}:(1,1,0). \\ 
\end{array}%
\end{align}
where $\nu _{L}^{i}$ and $e_{L}^{i}$ ($e_{L},\mu _{L},\tau _{L}$) are the
neutral and charged lepton families, respectively. Let's note that we assign
the right-handed leptons as $SU(3)_{L}$ singlets, which implies that their $%
U(1)_{X}$ quantum numbers correspond to their electric charges. The exotic
leptons of the model are: three neutral Majorana leptons $(\nu
^{1,2,3})_{L}^{c}$ and three right-handed Majorana leptons $N_{R}^{1,2,3}$
(A recent discussion of double and inverse see-saw neutrino mass generation
mechanisms in the context of 3-3-1 models can be found in Ref. \cite%
{Catano:2012kw}).

\quad The scalar sector the 3-3-1 models includes: three $3$'s irreps of $%
SU(3)_{L}$, where one triplet $\chi $ gets a TeV scale vaccuum expectation
value (VEV) $v_{\chi }$, that breaks the $SU(3)_{L}\otimes U(1)_{X}$
symmetry down to $SU(2)_{L}\otimes U(1)_{Y}$, thus generating the masses of
non SM fermions and non SM gauge bosons; and two light triplets $\eta $ and $%
\rho $ acquiring electroweak scale VEVs $v_{\eta }$ and $v_{\rho }$,
respectively and thus providing masses for the fermions and gauge bosons of
the SM.

\quad Regarding the scalar sector of the minimal 331 model, we assign the
scalar fields in the following $[SU(3)_{L},U(1)_{X}]$ representations: 
\begin{align}
\chi & =%
\begin{pmatrix}
\chi _{1}^{0} \\ 
\chi _{2}^{-} \\ 
\frac{1}{\sqrt{2}}(\upsilon _{\chi }+\xi _{\chi }\pm i\zeta _{\chi }) \\ 
\end{pmatrix}%
:(3,-1/3),\hspace{1cm}\rho =%
\begin{pmatrix}
\rho _{1}^{+} \\ 
\frac{1}{\sqrt{2}}(\upsilon _{\rho }+\xi _{\rho }\pm i\zeta _{\rho }) \\ 
\rho _{3}^{+} \\ 
\end{pmatrix}%
:(3,2/3),  \notag \\
\eta & =%
\begin{pmatrix}
\frac{1}{\sqrt{2}}(\upsilon _{\eta }+\xi _{\eta }\pm i\zeta _{\eta }) \\ 
\eta _{2}^{-} \\ 
\eta _{3}^{0}%
\end{pmatrix}%
:(3,-1/3).  \label{331-scalar}
\end{align}

We extend the scalar sector of the minimal 331 model by adding the following
fifteen very heavy $SU(3)_{L}$ scalar singlets: 
\begin{eqnarray}
\sigma  &\sim &(1,0),\hspace{1cm}\phi :(1,0),\hspace{1cm}\zeta :(1,0),
\label{331scalarsextra} \\
\varphi _{j} &:&(1,0),\hspace{1cm}\xi _{j}:(1,0),  \notag \\
\tau _{j} &:&(1,0),\hspace{1cm}\Delta _{j}:(1,0),\hspace{1cm}j=1,2,\hspace{%
1cm}  \notag \\
\Sigma _{k} &:&(1,0),\hspace{1cm}k=1,2,3,4.  \notag
\end{eqnarray}%
We assign the scalars into $S_{3}$ doublet, and $S_{3}$ singlet
representions. The $S_{3}\otimes Z_{3}\otimes Z_{3}^{\prime }\otimes
Z_{8}\otimes Z_{16}$ assignments of the scalar fields are: 
\begin{eqnarray*}
\eta  &\sim &\left( \mathbf{1},e^{\frac{2\pi i}{3}},1,1,1\right) ,\hspace{1cm%
}\rho \sim \left( \mathbf{1,}e^{-\frac{2\pi i}{3}},1,1,1\right) , \\
\chi  &\sim &\left( \mathbf{1,}1,1,1,1\right) ,\hspace{1cm}\sigma \sim
\left( \mathbf{1}^{\prime }\mathbf{,}1,1,1,e^{-\frac{\pi i}{8}}\right)  \\
\phi  &\sim &\left( \mathbf{1}^{\prime }\mathbf{,}1,1,-i,1\right) ,\hspace{%
1cm}\zeta \sim \left( \mathbf{1}^{\prime }\mathbf{,}1,1,1,1\right) , \\
\xi  &\sim &\left( \mathbf{2,}1,1,-1,1\right) ,\hspace{1cm}\tau \sim \left( 
\mathbf{2,}1,1,i^{\frac{1}{2}},1\right) ,\hspace{1cm} \\
\varphi _{1} &\sim &\left( \mathbf{1,}e^{-\frac{2\pi i}{3}},1,-i,1\right) ,%
\hspace{1cm}\varphi _{2}\sim \left( \mathbf{1}^{\prime }\mathbf{,}e^{-\frac{%
2\pi i}{3}},1,-i,1\right) , \\
\Delta  &\sim &\left( \mathbf{2,}e^{-\frac{2\pi i}{3}},1,-i,1\right) ,%
\hspace{1cm}\Sigma _{1}\sim \left( \mathbf{1},1,e^{\frac{2\pi i}{3}},-1,e^{%
\frac{3i\pi }{8}}\right) , \\
\Sigma _{2} &\sim &\left( \mathbf{1},1,e^{\frac{2\pi i}{3}},-1,e^{\frac{%
2i\pi }{8}}\right) ,\hspace{1cm}\Sigma _{3}\sim \left( \mathbf{1}^{\prime
},1,e^{\frac{2\pi i}{3}},-1,e^{-\frac{i\pi }{7}}\right) , \\
\Sigma _{4} &\sim &\left( \mathbf{1},1,e^{-\frac{2\pi i}{3}},-1,1\right) .
\end{eqnarray*}

It has been shown in Ref. \cite{Hernandez:2015dga}, that the minimization
equations for the scalar potential involving the $S_{3}$ scalar doublet,
imply that the $S_{3}$ scalar doublets $\xi $, $\tau $ and $\Delta $ can
acquire the following VEV pattern: 
\begin{equation}
\left\langle \xi \right\rangle =v_{\xi }\left( 1,0\right) ,\hspace{1cm}%
\left\langle \tau \right\rangle =v_{\tau }\left( 1,1\right) ,\hspace{1cm}%
\left\langle \Delta \right\rangle =v_{\Delta }\left( 1,0\right) .
\label{VEV}
\end{equation}%
The vacuum configuration of a $S_{3}$ scalar doublet, pointing either in the 
$\left( 1,0\right) $ or in the $\left( 1,1\right) $ $S_{3}$ directions, has
been considered in several $S_{3}$ flavor models (see for instance Refs. 
\cite{Kubo:2004ps,Hernandez:2014vta,Hernandez:2015dga}). In our model we
assume the hierarchy $v_{\Delta }<<v_{\tau }<<v_{\xi }$, between the VEVs of
the $S_{3}$ scalar doublets in order to neglect the mixings between these
fields and to treat their scalar potentials independently. Let us note that
mixing angles between those fields are suppressed by the ratios of their
VEVs, as follows from the method of recursive expansion of Ref. \cite%
{Grimus:2000vj}.

In the concerning to the lepton sector, we have the following $S_{3}\otimes
Z_{3}\otimes Z_{3}^{\prime }\otimes Z_{8}\otimes Z_{16}$ assignments: 
\begin{eqnarray}
L_{L}^{1} &\sim &\left( \mathbf{1,}e^{\frac{2\pi i}{3}},1,i^{\frac{1}{2}%
},1\right) ,\hspace{1cm}L_{L}=\left( L_{L}^{2},L_{L}^{3}\right) \sim \left( 
\mathbf{2,}e^{\frac{2\pi i}{3}},1,i^{\frac{1}{2}},1\right)  \notag \\
e_{R} &\sim &\left( \mathbf{1}^{\prime }\mathbf{,}e^{-\frac{2\pi i}{3}},1,i^{%
\frac{1}{2}},-1\right) ,\hspace{1cm}\mu _{R}\sim \left( \mathbf{1}^{\prime }%
\mathbf{,}e^{-\frac{2\pi i}{3}},1,1,e^{\frac{\pi i}{4}}\right) ,\hspace{1cm}
\notag \\
\tau _{R} &\sim &\left( \mathbf{1}^{\prime }\mathbf{,}e^{-\frac{2\pi i}{3}%
},1,1,1\right) \hspace{1cm}N_{R}^{1}\sim \left( \mathbf{1,}e^{\frac{2\pi i}{3%
}},1,i^{\frac{1}{2}},1\right)  \notag \\
N_{R} &=&\left( N_{R}^{2},N_{R}^{3}\right) \sim \left( \mathbf{2,}e^{\frac{%
2\pi i}{3}},1,i^{\frac{1}{2}},1\right) ,
\end{eqnarray}%
while the $S_{3}\otimes Z_{3}\otimes Z_{3}^{\prime }\otimes Z_{8}\otimes
Z_{16}$ assignments for the quark sector are: 
\begin{eqnarray}
Q_{L} &=&\left( Q_{1L},Q_{2L}\right) \sim \left( \mathbf{2},1,1,-1,e^{-\frac{%
i\pi }{8}}\right) ,\hspace{1cm}Q_{L}^{3}\sim \left( \mathbf{1,}%
1,1,1,1\right) ,  \notag \\
U_{R}^{1} &\sim &\left( \mathbf{1,}e^{-\frac{2\pi i}{3}},e^{\frac{2\pi i}{3}%
},1,e^{\frac{6i\pi }{8}}\right) ,\hspace{1cm}U_{R}^{2}\sim \left( \mathbf{1}%
^{\prime }\mathbf{,}e^{-\frac{2\pi i}{3}},e^{\frac{2\pi i}{3}},1,e^{\frac{%
2i\pi }{8}}\right) ,\hspace{1cm}U_{R}^{2}\sim \left( \mathbf{1,}e^{-\frac{%
2\pi i}{3}},1,1,1\right) ,  \notag \\
D_{R}^{1} &\sim &\left( \mathbf{1},e^{-\frac{2\pi i}{3}},1,1,e^{\frac{5i\pi 
}{8}}\right) ,\hspace{1.5cm}D_{R}^{2}\sim \left( \mathbf{1},e^{-\frac{2\pi i%
}{3}},e^{-\frac{2\pi i}{3}},-1,e^{\frac{3i\pi }{8}}\right) ,\hspace{1.5cm}%
D_{R}^{3}\sim \left( \mathbf{1}^{\prime },e^{-\frac{2\pi i}{3}%
},1,-1,1\right) ,  \notag \\
T_{R} &\sim &\left( \mathbf{1}^{\prime }\mathbf{,}1,1,1,1\right) ,\hspace{1cm%
}J_{R}^{1}\sim \left( \mathbf{1}^{\prime }\mathbf{,}1,1,1,-1\right) ,\hspace{%
1cm}J_{R}^{2}\sim \left( \mathbf{1}^{\prime }\mathbf{,}1,1,1,-i\right) .
\end{eqnarray}%
In the following we explain the role each discrete group factors of our
model. The $S_{3}$, $Z_{3}$, $Z_{3}^{\prime }$ and $Z_{8}$ discrete groups
reduce the number of the $SU(3)_{C}\otimes SU(3)_{L}\otimes U(1)_{X}$ model
parameters. This allow us to get viable textures for the fermion sector that
successfully describe the prevailing pattern of fermion masses and mixings,
as we will show in sections \ref{leptonmassesandmixing} and \ref%
{quarkmassesandmixing}. Let us note that we use the $S_{3}$ discrete group
since it is the smallest non-Abelian group that has been considerably
studied in the literature. It is worth mentioning that the $SU(3)_{L}$
scalar triplets are assigned to a $S_{3}$ trivial singlet representation,
whereas the $SU(3)_{L}$ scalar singlets are accomodated into three $S_{3}$
doublets, three $S_{3}$ trivial singlets and three $S_{3}$ non trivial
singlets. The $Z_{3}$ and $Z_{8}$ symmetries determines the allowed entries
of the charged lepton mass matrix. Furthermore, the $Z_{3}$ symmetry
distinguishes the right handed exotic quaks, being neutral under $Z_{3}$
from the right handed SM quarks, charged under this symmetry. Note that SM
right handed quarks are the only quark fields transforming non trivially
under the $Z_{3}$ symmetry. This results in the absence of mixing between SM
quarks and exotic quarks. Consequently, the $Z_{3}$ symmetry is crucial for
decoupling the SM quarks from the exotic quarks. Besides that, the $%
Z_{3}^{\prime }$ symmetry selects the allowed entries of the SM quark mass
matrices. Besides that, the $Z_{8}$ symmetry separates the $S_{3}$ scalar
doublets participating in the quark Yukawa interactions from those ones
participating in the charged lepton and neutrino Yukawa interactions. The $%
Z_{16}$ symmetry generates the hierarchy among charged fermion masses and
quark mixing angles that yields the observed charged fermion mass and quark
mixing pattern. It is worth mentioning that the properties of the $Z_{N}$
groups imply that the $Z_{16}$ symmetry is the smallest cyclic symmetry that
allows to build the Yukawa term $\overline{L}_{L}^{1}\rho e_{R}\frac{\sigma
^{8}}{\Lambda ^{8}}$ of dimension twelve from a $\frac{\sigma ^{8}}{\Lambda
^{8}}$ insertion on the $\overline{L}_{L}^{1}\rho e_{R}$ operator, crucial
to get the required $\lambda ^{8}$ suppression (where $\lambda =0.225$ is
one of the Wolfenstein parameters) needed to naturally explain the smallness
of the electron mass.

\quad Now let us briefly comment about a posible large discrete symmetry
group that could be used to embed the $S_{3}\otimes Z_{3}\otimes
Z_{3}^{\prime }\otimes Z_{8}\otimes Z_{16}$ discrete symmetry of our model.
Considering that the discrete group $\Delta \left( 6N^{2}\right) $ is
isomorphic to$\ (Z_{N}\times Z_{N}^{\prime })\rtimes S_{3}$ \cite%
{Ishimori:2010au} and the fact the $Z_{24}$ discrete group is the smallest
cyclic group that contains the $Z_{3}$ and $Z_{8}$ symmetries and the $%
Z_{3}^{\prime }$ symmetry is contained in the $Z_{24}^{\prime }$ group, it
follows that the $S_{3}\otimes Z_{3}\otimes Z_{3}^{\prime }\otimes
Z_{8}\otimes Z_{16}$ discrete group of our model can be embedded in the $%
\Delta \left( 6N^{2}\right) =\Delta $ $\left( 3456\right) $ discrete group
(where $N=24$). It would be interesting to implement the $\Delta \left(
6N^{2}\right) $ discrete symmetry in the 331 model and to study its
implications on fermion masses and mixings. This requieres careful studies
that are left beyond the scope of the present paper and will be done
elsewhere.

\quad With the aforementioned field content of our model, the relevant quark
and lepton Yukawa terms invariant under the group $\mathcal{G}$, take the
form:

\begin{eqnarray}
\tciLaplace _{Y}^{\left( Q\right) } &=&y_{33}^{\left( U\right) }\overline{Q}%
_{L}^{3}\eta U_{R}^{3}+y_{23}^{\left( U\right) }\overline{Q}_{L}^{2}\rho
^{\ast }U_{R}^{3}\frac{\xi \sigma }{\Lambda ^{2}}+y_{22}^{\left( U\right) }%
\overline{Q}_{L}^{2}\rho ^{\ast }U_{R}\frac{\xi \sigma ^{3}}{\Lambda ^{4}}%
+y_{11}^{\left( U\right) }\overline{Q}_{L}^{1}\rho ^{\ast }U_{R}\frac{\xi
\sigma ^{7}}{\Lambda ^{8}}  \notag \\
&&+y_{33}^{\left( D\right) }\overline{Q}_{L}^{3}\rho D_{R}^{3}\frac{\sigma
^{2}\Sigma _{2}}{\Lambda ^{3}}+y_{22}^{\left( D\right) }\overline{Q}_{L}\eta
^{\ast }D_{R}^{2}\frac{\xi \Sigma _{3}\sigma ^{3}}{\Lambda ^{5}}%
+y_{12}^{\left( D\right) }\overline{Q}_{L}\eta ^{\ast }D_{R}^{2}\frac{\xi
\Sigma _{4}\sigma ^{4}}{\Lambda ^{6}}  \notag \\
&&+y_{13}^{\left( D\right) }\overline{Q}_{L}\eta ^{\ast }D_{R}^{3}\frac{\xi
\sigma ^{4}\Sigma _{1}}{\Lambda ^{6}}+y_{11}^{\left( D\right) }\overline{Q}%
_{L}\eta ^{\ast }D_{R}^{1}\frac{\xi \sigma ^{6}}{\Lambda ^{7}}  \notag \\
&&+y^{\left( T\right) }\overline{Q}_{L}^{3}\chi T_{R}+y_{1}^{\left( J\right)
}\overline{Q}_{L}^{1}\chi ^{\ast }J_{R}^{1}+y_{2}^{\left( J\right) }%
\overline{Q}_{L}^{2}\chi ^{\ast }J_{R}^{2}+H.c  \label{YukawatermsQ}
\end{eqnarray}%
\begin{eqnarray}
-\mathcal{L}_{Y}^{\left( L\right) } &=&h_{1\rho e}^{\left( L\right) }%
\overline{L}_{L}^{1}\rho e_{R}\frac{\sigma ^{8}}{\Lambda ^{8}}+h_{1\rho \mu
}^{\left( L\right) }\left( \overline{L}_{L}\rho \tau \right) _{\mathbf{%
\mathbf{1}}}\mu _{R}\frac{\sigma ^{2}}{\Lambda ^{3}}+h_{2\rho \mu }^{\left(
L\right) }\left( \overline{L}_{L}\rho \tau \right) _{\mathbf{1}^{\prime
}}\mu _{R}\frac{\sigma ^{2}\zeta }{\Lambda ^{4}}  \notag \\
&&+h_{1\rho \tau }^{\left( L\right) }\left( \overline{L}_{L}\rho \tau
\right) _{\mathbf{1}}\tau _{R}\frac{1}{\Lambda }+h_{2\rho \tau }^{\left(
L\right) }\left( \overline{L}_{L}\rho \tau \right) _{\mathbf{1}^{\prime
}}\tau _{R}\frac{\zeta }{\Lambda ^{2}}+h_{1\chi }^{\left( L\right) }\left( 
\overline{L}_{L}\chi N_{R}\right) _{\mathbf{\mathbf{1}}}+h_{3\chi }^{\left(
L\right) }\overline{L}_{L}^{1}\chi N_{R}^{1}  \notag \\
&&+\frac{1}{2}h_{1N}\left( \overline{N}_{R}N_{R}^{C}\right) _{\mathbf{1}%
}\varphi _{1}+\frac{1}{2}h_{2N}\overline{N}_{R}^{1}N_{R}^{1}\varphi _{1}+%
\frac{1}{2}h_{3N}\left( \overline{N}_{R}N_{R}^{C}\right) _{\mathbf{1}%
^{\prime }}\varphi _{2}  \notag \\
&&+h_{\rho }^{\left( 1\right) }\varepsilon _{abc}\left( \overline{L}%
_{L}^{a}\left( L_{L}^{C}\right) ^{b}\right) _{\mathbf{1}^{\prime }}\left(
\rho ^{\ast }\right) ^{c}\frac{\phi }{\Lambda }+h_{\rho }^{\left( 2\right)
}\varepsilon _{abc}\left( \overline{L}_{L}^{a}\left( L_{L}^{1C}\right)
^{b}\left( \rho ^{\ast }\right) ^{c}\Delta \right) _{\mathbf{1}}\frac{1}{%
\Lambda }  \notag \\
&&+h_{\rho }^{\left( 3\right) }\varepsilon _{abc}\left( \left( \overline{L}%
_{L}^{1}\right) ^{a}\left( L_{L}^{C}\right) ^{b}\left( \rho ^{\ast }\right)
^{c}\Delta \right) _{\mathbf{1}}\frac{1}{\Lambda }+H.c,  \label{YukawatermsL}
\end{eqnarray}%
where the dimensionless couplings in Eqs. (\ref{YukawatermsQ}) and (\ref%
{YukawatermsL}) are $\mathcal{O}(1)$ parameters.

\quad Considering that the charged fermion mass and quark mixing pattern
arises from the breaking of the $Z_{3}\otimes Z_{3}^{\prime }\otimes
Z_{8}\otimes Z_{16}$ discrete group, we set the VEVs of the $SU(3)_{L}$
singlet scalars $\sigma $, $\zeta $, $\phi $, $\tau _{j}$, $\Delta _{j}$, $%
\xi _{j}$ ($j=1,2$) and $\Sigma _{k}$ ($k=1,2,3,4$)\ scalar singlets, as
follows:

\begin{equation}
v_{\phi }\sim v_{\Delta }\sim \lambda ^{5}\Lambda <<v_{\tau }=\lambda
^{3}\Lambda <<v_{\sigma }=v_{\zeta }=v_{\xi }=v_{\Sigma _{k}}=\Lambda
_{int}=\lambda \Lambda ,\hspace{1cm}k=1,2,3,4,  \label{VEVsinglets}
\end{equation}%
where $\lambda =0.225$ is one of the parameters of the Wolfenstein
parametrization and $\Lambda $ the cutoff of our model. Let us note that the 
$SU(3)_{L}$ singlet scalar fields $\sigma $, $\zeta $, $\xi _{j}$ ($j=1,2$)
and $\Sigma _{k}$ ($k=1,2,3,4$) having the VEVs of the same order of
magnitude are the ones that appear in the SM charged fermion Yukawa terms,
thus playing an important role in generating the SM charged fermion masses
and quark mixing angles. Regarding the $SU(3)_{L}$ singlet scalar fields $%
\tau _{j}$ ($j=1,2$), that participates in the charged lepton Yukawa
interactions, we assume that it acquires a VEV, much smaller than $\lambda
\Lambda $\ (we set its VEV as $\lambda ^{3}\Lambda $) in order to supress
its mixing with the $S_{3}$ scalar doublet $\xi $, which allows us to treat
their scalar potentials independently. Because of the reason mentioned
above, we have also assumed that the $S_{3}$ scalar doublet $\Delta $, that
appears in the Dirac neutrino Yukawa terms, acquires a VEV much smaller than 
$\lambda ^{3}\Lambda $, which we set close to $\lambda ^{5}\Lambda $.
Furthermore, in order to have the neutrino sector model parameters of the
same order, we have assumed that $v_{\phi }\sim v_{\Delta }$. As previously
mentioned, that aforemetioned hierarchy between the VEVs of the $S_{3}$
scalar doublets $\xi $, $\tau $ and $\Delta $ allows us to treat their
scalar potentials independently, thus providing a more natural justification
for their chosen VEV patterns given in Eq. (\ref{VEV}) as natural solutions
of the scalar potential minimization equations for the whole region of
parameter space.

As we will explain in the following, we are going to implement an inverse
seesaw mechanism for the generation of the light active neutrino masses. To
implement an inverse seesaw mechanism, we need very light right handed
Majorana neutrinos, which implies that the $SU(3)_{L}$ singlet scalars
having Yukawa interactions with those neutrinos should acquire very small
vacuum expectation values, much smaller than the scale of breaking of the SM
electroweak symmetry. Because of this reason, we further assume that the $%
SU(3)_{L}$ scalar singlets $\varphi _{j}$ ($j=1,2$) giving masses to the
right handed Majorana neutrinos have VEVs much smaller than the electroweak
symmetry breaking scale, then providing small masses to these Majorana
neutrinos, and thus giving rise to a inverse seesaw mechanism of active
neutrino masses. Therefore, we have the following hierarchy among the VEVs
of the scalar fields in our model: 
\begin{equation}
v_{\varphi _{1}}\sim v_{\varphi _{2}}<<v_{\rho }\sim v_{\eta }\sim
v<<v_{\chi }<<v_{\phi }\sim v_{\Delta }<<v_{\tau }<<\Lambda _{int}.
\label{VEVhierarchy}
\end{equation}

In summary, for the reasons mentioned above and considering a very high
model cutoff $\Lambda \gg v_{\chi }$, we set the vacuum expectation values
(VEVs) of the $SU(3)_{L}$ scalar singlets at a very high energy, much larger
than $v_{\chi }\approx \mathcal{O}(1)$ TeV, with the exception of the VEVs
of $\varphi _{j}$, $\Delta _{j}$ ($j=1,2$), taken to be much smaller than
the electroweak symmetry breaking scale $v=246$ GeV. It is noteworthy the $%
SU(3)_{C}\otimes SU(3)_{L}\otimes U(1)_{X}\otimes Z_{3}\otimes Z_{3}^{\prime
}\otimes Z_{8}\otimes Z_{16}$ symmetry is broken down to $SU(3)_{C}\otimes
SU(3)_{L}\otimes U(1)_{X}\otimes Z_{3}$, at the scale $\Lambda _{int}$, by
the vacuum expectation values of the $SU(3)_{L}$ singlet scalar fields $%
\sigma $, $\zeta $, $\xi _{j}$ and $\Sigma _{k}$ ($k=1,2,3,4$).

It is worth mentioning that in order that the small VEVs of the $SU(3)_{L}$
scalar singlets $\varphi _{j}$ ($j=1,2$) be stable under radiative
corrections, a Veltmann condition that connects a combination of the quartic
couplings of the scalar potential that involve a pair of these scalar fields
with the remaining ones and the combination of the Yukawa couplings of these
scalar singlets with the right handed Majorana neutrinos, has to be
fullfilled. That Veltmann condition will arise by requiring the cancellation
of the quadratically divergent scalar and fermionic contributions,
contributions that interfere destructively. The aforementioned Veltmann
condition will constrain the quartic scalar couplings of the scalar
interactions involving a pair of the scalar fields that acquire very small
VEVs. The resulting constraints on these quartic scalar couplings will not
affect neither the fermions masses and mixings nor the flavor changing top
quark decays. Having the VEVs of the scalar fields of our model stable under
radiative corrections in the whole region of parameter space, will require
to embed our model in a warped five dimensional framework or to implement
supersymmetry. This requires careful studies which are left beyond the scope
of the present paper.

\subsection{Low energy scalar potential}

\label{scalarpotential} The renormalizable low energy scalar potential of
the model takes the form:%
%
%
%
%
%
%
%
%
%
%
%
%
%
%
%
%
%
%
%
%
%
%
%
%
%
%
%
%
%
\begin{eqnarray}
&&V_{H}=\mu _{\chi }^{2}(\chi ^{\dagger }\chi )+\mu _{\eta }^{2}(\eta
^{\dagger }\eta )+\mu _{\rho }^{2}(\rho ^{\dagger }\rho )+f\left( \eta
_{i}\chi _{j}\rho _{k}\varepsilon ^{ijk}+H.c.\right) +\lambda _{1}(\chi
^{\dagger }\chi )(\chi ^{\dagger }\chi )  \notag \\
&&+\lambda _{2}(\rho ^{\dagger }\rho )(\rho ^{\dagger }\rho )+\lambda
_{3}(\eta ^{\dagger }\eta )(\eta ^{\dagger }\eta )+\lambda _{4}(\chi
^{\dagger }\chi )(\rho ^{\dagger }\rho )+\lambda _{5}(\chi ^{\dagger }\chi
)(\eta ^{\dagger }\eta )  \notag \\
&&+\lambda _{6}(\rho ^{\dagger }\rho )(\eta ^{\dagger }\eta )+\lambda
_{7}(\chi ^{\dagger }\eta )(\eta ^{\dagger }\chi )+\lambda _{8}(\chi
^{\dagger }\rho )(\rho ^{\dagger }\chi )+\lambda _{9}(\rho ^{\dagger }\eta
)(\eta ^{\dagger }\rho ).  \label{v00}
\end{eqnarray}

After the symmetry breaking, it is found that the scalar mass eigenstates
are connected with the weak scalar states by the following relations 
\begin{eqnarray}
\begin{pmatrix}
G_{1}^{\pm } \\ 
H_{1}^{\pm } \\ 
\end{pmatrix}%
=R_{\beta _{T}}%
\begin{pmatrix}
\rho _{1}^{\pm } \\ 
\eta _{2}^{\pm } \\ 
\end{pmatrix}
&,&\hspace{0.3cm}%
\begin{pmatrix}
G_{1}^{0} \\ 
A_{1}^{0} \\ 
\end{pmatrix}%
=R_{\beta _{T}}%
\begin{pmatrix}
\zeta _{\rho } \\ 
\zeta _{\eta } \\ 
\end{pmatrix}%
,\hspace{0.3cm}%
\begin{pmatrix}
H_{1}^{0} \\ 
h^{0} \\ 
\end{pmatrix}%
=R_{\alpha _{T}}%
\begin{pmatrix}
\xi _{\rho } \\ 
\xi _{\eta } \\ 
\end{pmatrix}%
,  \label{331-mass-scalar-a} \\
\begin{pmatrix}
G_{2}^{0} \\ 
H_{2}^{0} \\ 
\end{pmatrix}%
=R%
\begin{pmatrix}
\chi _{1}^{0} \\ 
\eta _{3}^{0} \\ 
\end{pmatrix}
&,&\hspace{0.3cm}%
\begin{pmatrix}
G_{3}^{0} \\ 
H_{3}^{0} \\ 
\end{pmatrix}%
=R%
\begin{pmatrix}
\zeta _{\chi } \\ 
\xi _{\chi } \\ 
\end{pmatrix}%
,\hspace{0.3cm}%
\begin{pmatrix}
G_{2}^{\pm } \\ 
H_{2}^{\pm } \\ 
\end{pmatrix}%
=R%
\begin{pmatrix}
\chi _{2}^{\pm } \\ 
\rho _{3}^{\pm } \\ 
\end{pmatrix}%
,  \label{331-mass-scalar-b}
\end{eqnarray}

with

\begin{equation}
R_{\alpha _{T}(\beta _{T})}=\left( 
\begin{array}{cc}
\cos \alpha _{T}(\beta _{T}) & \sin \alpha _{T}(\beta _{T}) \\ 
-\sin \alpha _{T}(\beta _{T}) & \cos \alpha _{T}(\beta _{T})%
\end{array}%
\right) ,\hspace{2cm}R=\left( 
\begin{array}{cc}
-1 & 0 \\ 
0 & 1%
\end{array}%
\right) ,
\end{equation}%
where $\tan \beta _{T}=v_{\eta }/v_{\rho }$, and $\tan 2\alpha
_{T}=M_{1}^{2}/(M_{2}^{2}-M_{3}^{2})$ with: 
\begin{eqnarray}
M_{1}^{2} &=&4\lambda _{6}v_{\eta }v_{\rho }+2\sqrt{2}fv_{\chi },  \notag \\
M_{2}^{2} &=&4\lambda _{2}v_{\rho }^{2}-\sqrt{2}fv_{\chi }\tan \beta _{T}, 
\notag \\
M_{3}^{2} &=&4\lambda _{3}v_{\eta }^{2}-\sqrt{2}fv_{\chi }/\tan \beta _{T}.
\end{eqnarray}

The low energy physical scalar spectrum of our model includes: 4 massive
charged Higgs ($H_{1}^{\pm }$, $H_{2}^{\pm }$), one CP-odd Higgs ($A_{1}^{0}$%
), 3 neutral CP-even Higgs ($h^{0},H_{1}^{0},H_{3}^{0}$) and 2 neutral Higgs
($H_{2}^{0},\overline{H}_{2}^{0}$) bosons. The scalar $h^{0}$ is identified
with the SM-like $126$ GeV Higgs boson found at the LHC. It it noteworthy
that the neutral Goldstone bosons $G_{1}^{0}$, $G_{3}^{0}$, $G_{2}^{0}$ , $%
\overline{G}_{2}^{0}$ are associated to the longitudinal components of the $%
Z $, $Z^{\prime }$, $K^{0}$ and $\overline{K}^{0}$gauge bosons,
respectively. Furthermore, the charged Goldstone bosons $G_{1}^{\pm }$ and $%
G_{2}^{\pm }$ are associated to the longitudinal components of the $W^{\pm }$
and $K^{\pm } $ gauge bosons, respectively.

%
%

\section{Lepton masses and mixings}

\label{leptonmassesandmixing} From Eqs. (\ref{VEV}), (\ref{YukawatermsL}), (%
\ref{VEVsinglets}) and using the product rules of the $S_{3}$ group given in
Appendix \ref{S3}, it follows that the mass matrix for charged leptons is: 

\begin{equation}
M_{l}=\left( 
\begin{array}{ccc}
a_{11}^{\left( l\right) }\lambda ^{8} & 0 & 0 \\ 
0 & a_{22}^{\left( l\right) }\lambda ^{5} & a_{23}^{\left( l\right) }\lambda
^{3} \\ 
0 & a_{32}^{\left( l\right) }\lambda ^{5} & a_{33}^{\left( l\right) }\lambda
^{3}%
\end{array}%
\right) \frac{v}{\sqrt{2}}.  \label{Ml}
\end{equation}

Since the charged lepton mass hierarchy arises from the breaking of the $%
Z_{3}\otimes Z_{8}\otimes Z_{16}$ discrete group and in order to simplify
the analysis, we consider an scenario of approximate universality in the
dimensionless SM charged lepton Yukawa couplings, as follows: 
\begin{equation}
a_{32}^{\left( l\right) }=a_{4}^{\left( l\right) },\hspace{1cm}%
a_{23}^{\left( l\right) }=a_{4}^{\left( l\right) }e^{-i\alpha }
\end{equation}

where $a_{11}^{\left( l\right) }$, $a_{22}^{\left( l\right) }$, $%
a_{33}^{\left( l\right) }$ and $a_{4}^{\left( l\right) }$ are assumed to be
real $\mathcal{O}(1)$ parameters.

The matrix $M_{l}M_{l}^{\dagger }$ is diagonalized by a rotation matrix $%
R_{l}$ according to: 
\begin{eqnarray}
R_{l}^{\dagger }M_{l}M_{l}^{\dagger }R_{l} &=&diag\left( m_{e},m_{\mu
},m_{\tau }\right) ,\hspace{1cm}R_{l}=\left( 
\begin{array}{ccc}
1 & 0 & 0 \\ 
0 & \cos \theta _{l} & -\sin \theta _{l}e^{-i\alpha } \\ 
0 & \sin \theta _{l}e^{i\alpha } & \cos \theta _{l}%
\end{array}%
\right) ,  \notag \\
\tan \theta _{l} &\simeq &-\frac{a_{4}^{\left( l\right) }}{a_{33}^{\left(
l\right) }},\hspace{1cm}\cos \theta _{l}\simeq \frac{a_{33}^{\left( l\right)
}}{\sqrt{\left( a_{33}^{\left( l\right) }\right) ^{2}+\left( a_{4}^{\left(
l\right) }\right) ^{2}}},\hspace{1cm}\sin \theta _{l}\simeq -\frac{%
a_{4}^{\left( l\right) }}{\sqrt{\left( a_{33}^{\left( l\right) }\right)
^{2}+\left( a_{4}^{\left( l\right) }\right) ^{2}}},  \label{Rl}
\end{eqnarray}

where, from Eq. (\ref{Ml}) it follows that the charged lepton masses are
approximatelly given by:

\begin{equation}
m_{e}=a_{11}^{\left( l\right) }\lambda ^{8}\frac{v}{\sqrt{2}},\hspace{1cm}%
m_{\mu }\simeq \frac{\left\vert a_{22}^{\left( l\right) }a_{33}^{\left(
l\right) }-\left( a_{4}^{\left( l\right) }\right) ^{2}\right\vert }{\sqrt{%
\left( a_{33}^{\left( l\right) }\right) ^{2}+\left( a_{4}^{\left( l\right)
}\right) ^{2}}}\lambda ^{5}\frac{v}{\sqrt{2}},\hspace{1cm}m_{\tau }\simeq 
\sqrt{\left( a_{33}^{\left( l\right) }\right) ^{2}+\left( a_{4}^{\left(
l\right) }\right) ^{2}}\lambda ^{3}\frac{v}{\sqrt{2}}.  \label{ml}
\end{equation}%
Note that the charged lepton masses are connected with the electroweak
symmetry breaking scale $v=246$ GeV by their scalings with powers of the
Wolfenstein parameter $\lambda =0.225$, with $\mathcal{O}(1)$ coefficients.
This is consistent with our previous assumption made in Eq. (\ref%
{VEVsinglets}) regarding the size of the VEVs for the $SU(3)_{L}$ singlet
scalars appearing in the charged fermion Yukawa terms. Furthermore, it is
noteworthy that the mixing angle $\theta _{l}$ in the charged lepton sector
is large, which gives rise to an important contribution to the leptonic
mixing matrix, coming from the mixing of charged leptons.

In the concerning to the neutrino sector, the following neutrino mass terms
arise: 
\begin{equation}
-\mathcal{L}_{mass}^{\left( \nu \right) }=\frac{1}{2}\left( 
\begin{array}{ccc}
\overline{\nu _{L}^{C}} & \overline{\nu _{R}} & \overline{N_{R}}%
\end{array}%
\right) M_{\nu }\left( 
\begin{array}{c}
\nu _{L} \\ 
\nu _{R}^{C} \\ 
N_{R}^{C}%
\end{array}%
\right) +H.c,  \label{Lnu}
\end{equation}%
where the $S_{3}$ discrete flavor group constrains the neutrino mass matrix
to be of the form:

\begin{eqnarray}
M_{\nu } &=&\left( 
\begin{array}{ccc}
0_{3\times 3} & M_{D} & 0_{3\times 3} \\ 
M_{D}^{T} & 0_{3\times 3} & M_{\chi } \\ 
0_{3\times 3} & M_{\chi }^{T} & M_{R}%
\end{array}%
\right) ,\hspace{0.7cm}M_{D}=\frac{v_{\xi }v_{\phi }v_{\rho }}{\sqrt{2}%
\Lambda ^{2}}\left( 
\begin{array}{ccc}
0 & a & 0 \\ 
-a & 0 & b \\ 
0 & -b & 0%
\end{array}%
\right) ,  \notag \\
M_{\chi } &=&h_{1\chi }^{\left( L\right) }\frac{v_{\chi }}{\sqrt{2}}\left( 
\begin{array}{ccc}
x & 0 & 0 \\ 
0 & 1 & 0 \\ 
0 & 0 & 1%
\end{array}%
\right) ,\hspace{0.7cm}M_{R}=h_{1N}v_{\varphi _{1}}\left( 
\begin{array}{ccc}
1 & 0 & 0 \\ 
0 & y & z \\ 
0 & z & y%
\end{array}%
\right) ,  \notag \\
a &=&h_{\rho }^{\left( 2\right) }-h_{\rho }^{\left( 3\right) },\hspace{0.7cm}%
b=2h_{\rho }^{\left( 1\right) },\hspace{0.7cm}x=\frac{h_{2\chi }^{\left(
L\right) }}{h_{1\chi }^{\left( L\right) }},\hspace{0.7cm}y=\frac{h_{2N}}{%
h_{1N}},\hspace{0.7cm}z=\frac{h_{3N}v_{\varphi _{2}}}{h_{1N}v_{\varphi _{1}}}%
.
\end{eqnarray}

Since the $SU(3)_{L}$ scalar singlets $\varphi _{j}$ ($j=1,2$) having Yukawa
interactions with the right handed Majorana neutrinos acquire VEVs much
smaller than the electroweak symmetry breaking scale, these Majorana
neutrinos are very light, so that the active neutrinos get small masses via
inverse seesaw mechanism.

As shown in detail in Ref. \cite{Catano:2012kw}, the full rotation matrix is
approximatelly given by: 
\begin{equation}
\mathbb{U}=%
\begin{pmatrix}
V_{\nu } & B_{3}U_{\chi } & B_{2}U_{R} \\ 
-\frac{(B_{2}^{\dagger }+B_{3}^{\dagger })}{\sqrt{2}}V_{\nu } & \frac{(1-S)}{%
\sqrt{2}}U_{\chi } & \frac{(1+S)}{\sqrt{2}}U_{R} \\ 
-\frac{(B_{2}^{\dagger }-B_{3}^{\dagger })}{\sqrt{2}}V_{\nu } & \frac{(-1-S)%
}{\sqrt{2}}U_{\chi } & \frac{(1-S)}{\sqrt{2}}U_{R}%
\end{pmatrix}%
,  \label{U}
\end{equation}%
where 
\begin{equation}
S=-\frac{1}{2\sqrt{2}h_{\chi }^{\left( L\right) }v_{\chi }}M_{R},\hspace{1cm}%
\hspace{1cm}B_{2}\simeq B_{3}\simeq \frac{1}{h_{\chi }^{\left( L\right)
}v_{\chi }}M_{D}^{\ast },
\end{equation}%
and the physical neutrino mass matrices are: 
\begin{eqnarray}
M_{\nu }^{\left( 1\right) } &=&M_{D}\left( M_{\chi }^{T}\right)
^{-1}M_{R}M_{\chi }^{-1}M_{D}^{T},  \label{Mnu1} \\
M_{\nu }^{\left( 2\right) } &=&-M_{\chi }^{T}+\frac{1}{2}M_{R},\hspace{1cm}%
\hspace{1cm}M_{\nu }^{\left( 3\right) }=M_{\chi }^{T}+\frac{1}{2}M_{R},
\label{Mnu2}
\end{eqnarray}%
where $M_{\nu }^{\left( 1\right) }$ corresponds to the active neutrino mass
matrix whereas $M_{\nu }^{\left( 2\right) }$ and $M_{\nu }^{\left( 3\right)
} $ are the exotic Dirac neutrino mass matrices. Note that the physical
eigenstates include three active neutrinos and six exotic neutrinos. The
exotic neutrinos are pseudo-Dirac, with masses $\sim \pm M_{\chi }^{T}$ and
a small splitting $M_{R}$. Furthermore, $V_{\nu }$, $U_{R}$ and $U_{\chi }$
are the rotation matrices which diagonalize $M_{\nu }^{\left( 1\right) }$, $%
M_{\nu }^{\left( 2\right) }$ and $M_{\nu }^{\left( 3\right) }$, respectively.

Furthermore, as follows from Eq. (\ref{U}), we can connect the neutrino
fields $\nu _{L}=\left( \nu _{1L},\nu _{2L},\nu _{3L}\right) ^{T}$, $\nu
_{R}^{C}=\left( \nu _{1R}^{C},\nu _{2R}^{C},\nu _{3R}^{C}\right) $ and $%
N_{R}^{C}=\left( N_{1R}^{C},N_{2R}^{C},N_{3R}^{C}\right) $ with the neutrino
mass eigenstates by the following approximate relations: 
\begin{equation}
\left( 
\begin{array}{c}
\nu _{L} \\ 
\nu _{R}^{C} \\ 
N_{R}^{C}%
\end{array}%
\right) \simeq \left( 
\begin{array}{c}
V_{\nu }\xi _{L}^{\left( 1\right) } \\ 
\frac{1}{\sqrt{2}}U_{\chi }\xi _{L}^{\left( 2\right) }+\frac{1}{\sqrt{2}}%
U_{R}\xi _{L}^{\left( 3\right) } \\ 
-\frac{1}{\sqrt{2}}U_{\chi }\xi _{L}^{\left( 2\right) }+\frac{1}{\sqrt{2}}%
U_{R}\xi _{L}^{\left( 3\right) }%
\end{array}%
\right) ,\hspace{0.5cm}\hspace{0.5cm}\hspace{0.5cm}\hspace{0.5cm}\xi
_{L}^{\left( j\right) }=\left( 
\begin{array}{c}
\xi _{1L}^{\left( j\right) } \\ 
\xi _{2L}^{\left( j\right) } \\ 
\xi _{3L}^{\left( j\right) }%
\end{array}%
\right) ,\hspace{0.5cm}\hspace{0.5cm}\hspace{0.5cm}\hspace{0.5cm}j=1,2,3.
\end{equation}%
where $\xi _{kL}^{\left( 1\right) }$, $\xi _{kL}^{\left( 2\right) }$ and $%
\xi _{kL}^{\left( 3\right) }$ ($k=1,2,3$) are the three active neutrinos and
six exotic neutrinos, respectively.

\quad From Eq. (\ref{Mnu1}) it follows that the light active neutrino mass
matrix is given by: 
\begin{eqnarray}
M_{\nu }^{\left( 1\right) } &=&m_{\nu }\left( 
\begin{array}{ccc}
a^{2} & \kappa ab & -ab \\ 
\kappa ab & c^{2} & -\kappa b^{2} \\ 
-ab & -\kappa b^{2} & b^{2}%
\end{array}%
\right) ,  \notag \\
m_{\nu } &=&\frac{h_{1N}v_{\varphi _{1}}v_{\xi }^{2}v_{\phi }^{2}v_{\rho
}^{2}}{\left( h_{1\chi }^{\left( L\right) }\right) ^{2}v_{\chi }^{2}\Lambda
^{4}},\hspace{0.5cm}\hspace{0.5cm}\kappa =\frac{z}{y},  \notag \\
c^{2} &=&b^{2}+\frac{a^{2}}{x^{2}y}.  \label{Mnu}
\end{eqnarray}

Let us note that the smallness of the active neutrino masses arises from
their scaling with inverse powers of the high energy cutoff $\Lambda $ as
well as from their linear dependence on the very small VEVs of the $%
SU(3)_{L} $ singlets $\varphi _{j}$ ($j=1,2$), assumed to be of the same
order of magnitude.

Considering that the orders of magnitude of the SM particles and new physics
yield the constraints $v_{\chi }\gtrsim 1$ TeV and $v_{\eta }^{2}+v_{\rho
}^{2}=v^{2}$ and taking into account our assumption that the dimensionless
lepton Yukawa couplings are $\mathcal{O}(1)$ parameters, from Eq. (\ref{Mnu}%
) and the relations $v_{\xi }=\lambda \Lambda $, $v_{\phi }\sim \lambda
^{5}\Lambda $, $v_{\rho }\sim 100$ GeV, $v_{\chi }\sim 1$ TeV,\ we get that
the mass scale for the light active neutrinos satisfies $m_{\nu }\sim
10^{-10}v_{\varphi }$. Consequently, taking $m_{\nu }\sim 50$ meV, we find
for the VEV $v_{\varphi _{1}}$ of the singlet scalar $\varphi _{1}$, the
estimate 
\begin{equation}
v_{\varphi _{1}}\sim 0.5\text{ GeV}.  \label{cutoff}
\end{equation}

\begin{figure}[tbp]
\resizebox{18cm}{5cm}{\includegraphics{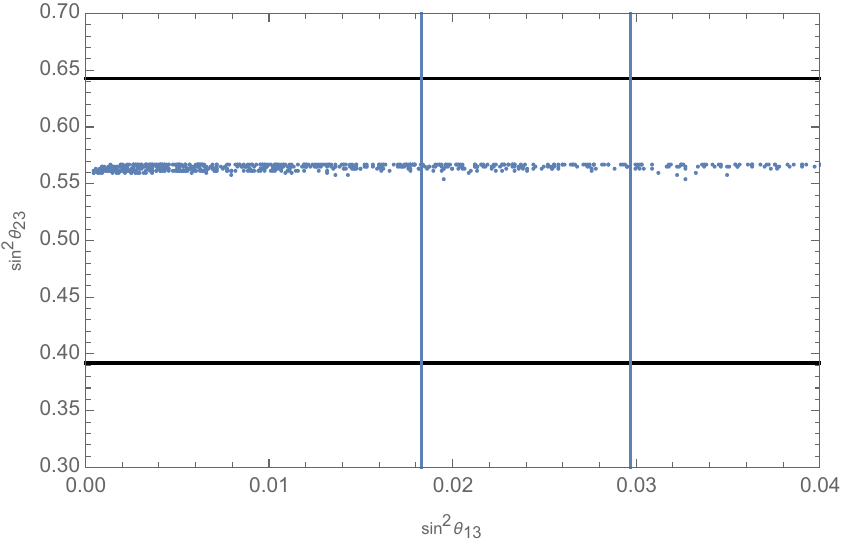}\includegraphics{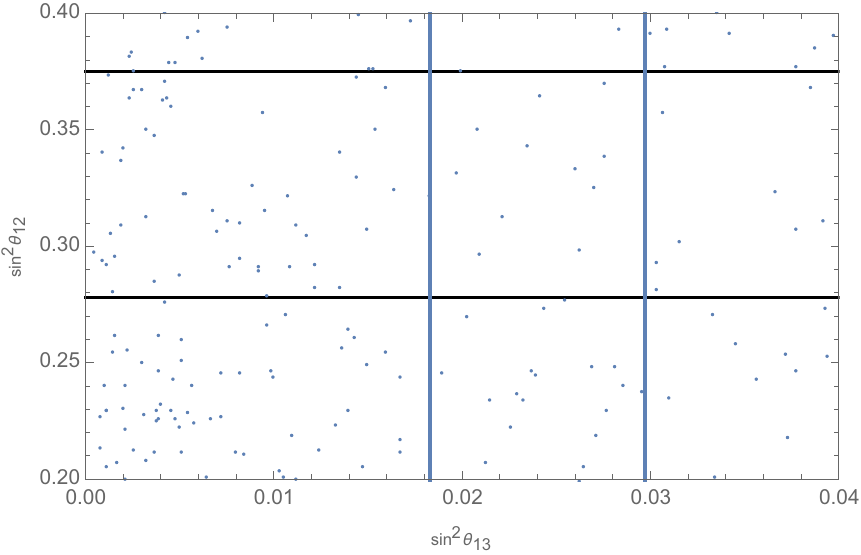}
\includegraphics{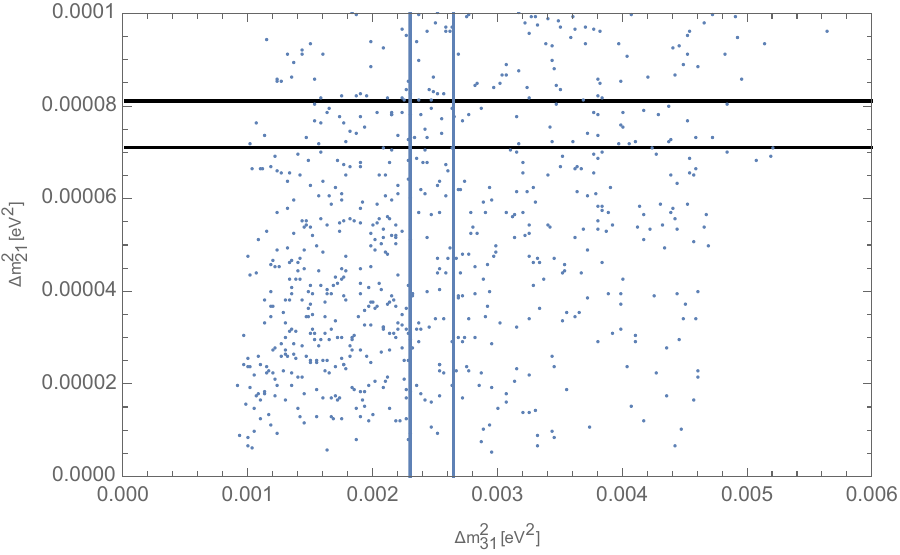}} 
\caption{Correlations between $\sin^2\protect\theta_{23}$ and $\sin^2\protect%
\theta_{13}$, $\sin^2\protect\theta_{12}$ and $\sin^2\protect\theta_{13}$, $%
\Delta m_{21}^{2}$ and $\Delta m_{31}^{2}$ for the case of normal hierarchy.
The horizonal and vertical lines are the minimum and maximum values of the
leptonic mixing parameters and neutrino mass squared splittings inside the $3%
\protect\sigma$ experimentally allowed range.}
\label{NH}
\end{figure}

\begin{figure}[tbp]
\resizebox{18cm}{5cm}{\includegraphics{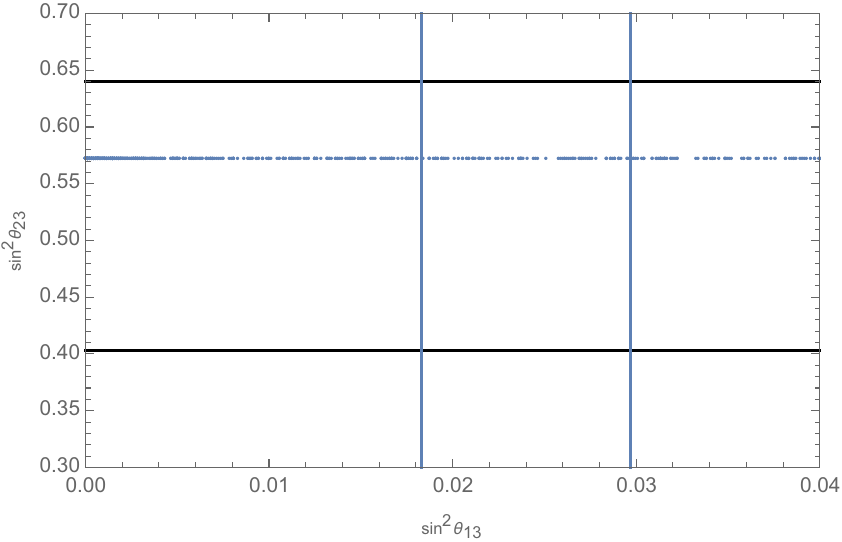}\includegraphics{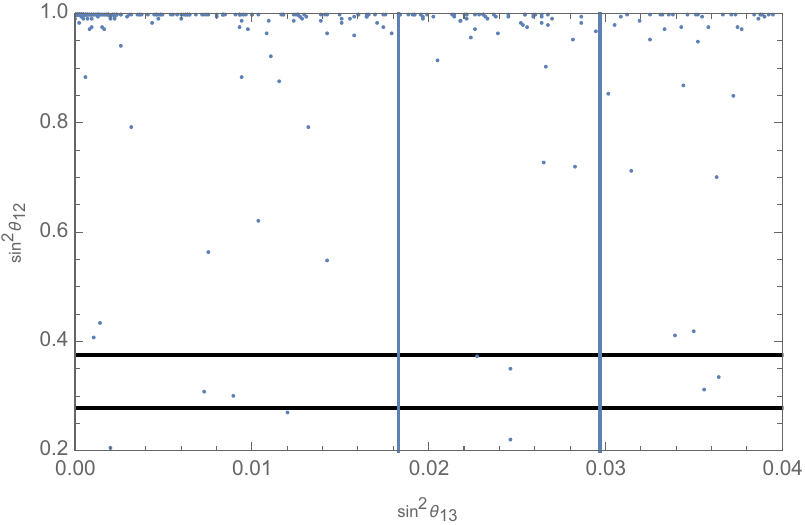}
\includegraphics{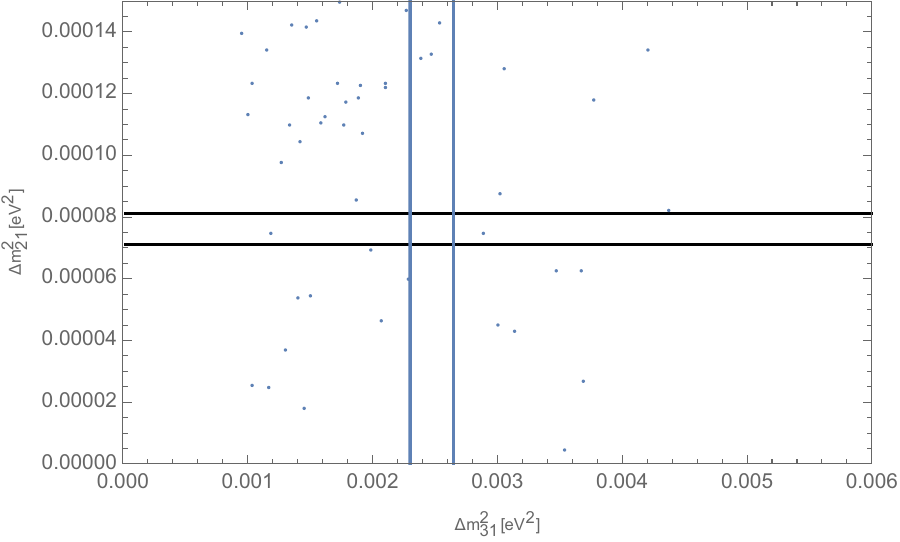}} 
\caption{Correlations between $\sin^2\protect\theta_{23}$ and $\sin^2\protect%
\theta_{13}$, $\sin^2\protect\theta_{12}$ and $\sin^2\protect\theta_{13}$, $%
\Delta m_{21}^{2}$ and $\Delta m_{31}^{2}$ for the case of inverted
hierarchy. The horizonal and vertical lines are the minimum and maximum
values of the leptonic mixing parameters and neutrino mass squared
splittings inside the $3\protect\sigma$ experimentally allowed range.}
\label{IH}
\end{figure}
In what follows we proceed to fit the lepton sector model parameters $m_{\nu
}$, $a_{11}^{\left( l\right) }$, $a_{22}^{\left( l\right) }$, $%
a_{33}^{\left( l\right) }$, $a_{4}^{\left( l\right) }$, $a$, $b$, $c$ and $%
\kappa $ to reproduce the experimental values for the physical observables
of the lepton sector, i.e., the {three charged lepton masses, the two
neutrino mass squared splittings and the three leptonic mixing angles. To
this end, we fix }$m_{\nu }=50$ meV and we vary the parameters $%
a_{11}^{\left( l\right) }$, $a_{22}^{\left( l\right) }$, $a_{33}^{\left(
l\right) }$, $a_{4}^{\left( l\right) }$, $a$, $b$, $c$ and $\kappa $ to fit
the charged lepton masses, the neutrino mass squared splitings $\Delta
m_{21}^{2}$, $\Delta m_{31}^{2}$\ (note that we define $\Delta
m_{ij}^{2}=m_{i}^{2}-m_{j}^{2}$) and the leptonic mixing angles $\sin
^{2}\theta _{12}$, $\sin ^{2}\theta _{13}$ and $\sin ^{2}\theta _{23}$\ to
their experimental values for normal (NH) and Inverted (IH) neutrino mass
hierarchy. The results shown in Table \ref{Observables0} correspond to the
following best-fit values:

\begin{eqnarray}
a_{11}^{\left( l\right) } &\simeq &0.42,\hspace{1cm}a_{22}^{\left( l\right)
}\simeq 1.88,\hspace{1cm}a_{33}^{\left( l\right) }\simeq 0.67,  \notag \\
a_{4}^{\left( l\right) } &\simeq &0.58,\hspace{1cm}a\simeq 0.28,\hspace{1cm}%
b\simeq 0.39  \notag \\
c &\simeq &-0.97,\hspace{1cm}\kappa \simeq 1.11  \notag \\
\theta _{l} &\simeq &-41.69^{\circ },\hspace{1cm}\alpha \simeq -85.99^{\circ
},\ \ \ \ \ \ \ \ \mbox{for NH}
\end{eqnarray}

\begin{eqnarray}
a_{11}^{\left( l\right) } &\simeq &0.42,\hspace{1cm}a_{22}^{\left( l\right)
}\simeq 2.33,\hspace{1cm}a_{33}^{\left( l\right) }\simeq 0.57,  \notag \\
a_{4}^{\left( l\right) } &\simeq &-0.67,\hspace{1cm}a\simeq 0.98,\hspace{1cm}%
b\simeq 0.15  \notag \\
c &\simeq &-0.99,\hspace{1cm}\kappa \simeq -0.05  \notag \\
\theta _{l} &\simeq &49.20^{\circ },\hspace{1cm}\alpha \simeq -93.60^{\circ
},\ \ \ \ \ \ \ \ \mbox{for IH}
\end{eqnarray}

\begin{table}[tbh]
\begin{center}
\begin{tabular}{c|l|l}
\hline\hline
Observable & Model value & Experimental value \\ \hline
$m_{e}(MeV)$ & \quad $0.487$ & \quad $0.487$ \\ \hline
$m_{\mu }(MeV)$ & \quad $102.8$ & \quad $102.8\pm 0.0003$ \\ \hline
$m_{\tau }(GeV)$ & \quad $1.75$ & \quad $1.75\pm 0.0003$ \\ \hline
$\Delta m_{21}^{2}$($10^{-5}$eV$^{2}$) (NH) & \quad $7.60$ & \quad $%
7.60_{-0.18}^{+0.19}$ \\ \hline
$\Delta m_{31}^{2}$($10^{-3}$eV$^{2}$) (NH) & \quad $2.48$ & \quad $%
2.48_{-0.07}^{+0.05}$ \\ \hline
$\sin ^{2}\theta _{12}$ (NH) & \quad $0.323$ & \quad $0.323\pm 0.016$ \\ 
\hline
$\sin ^{2}\theta _{23}$ (NH) & \quad $0.567$ & \quad $%
0.567_{-0.128}^{+0.032} $ \\ \hline
$\sin ^{2}\theta _{13}$ (NH) & \quad $0.0234$ & \quad $0.0234\pm 0.0020$ \\ 
\hline
$\delta $ (NH) & \quad $89.18^{\circ }$ & \quad Unknown \\ \hline
$\delta $ (IH) & \quad $86.40^{\circ }$ & \quad Unknown \\ \hline
$J$ (NH) & \quad $3.46\times 10^{-2}$ & \quad Unknown \\ \hline
$J$ (IH) & \quad $3.49\times 10^{-2}$ & \quad Unknown \\ \hline
$\Delta m_{21}^{2}$($10^{-5}$eV$^{2}$) (IH) & \quad $7.60$ & \quad $%
7.60_{-0.18}^{+0.19}$ \\ \hline
$\Delta m_{13}^{2}$($10^{-3}$eV$^{2}$) (IH) & \quad $2.38$ & \quad $%
2.38_{-0.06}^{+0.05}$ \\ \hline
$\sin ^{2}\theta _{12}$ (IH) & \quad $0.323$ & \quad $0.323\pm 0.016$ \\ 
\hline
$\sin ^{2}\theta _{23}$ (IH) & \quad $0.573$ & \quad $%
0.573_{-0.043}^{+0.025} $ \\ \hline
$\sin ^{2}\theta _{13}$ (IH) & \quad $0.0240$ & \quad $0.0240\pm 0.0019$ \\ 
\hline
\end{tabular}%
\end{center}
\par
\caption{Model and experimental values of the charged lepton masses,
neutrino mass squared splittings and leptonic mixing parameters for the
normal (NH) and inverted (IH) mass hierarchies. Model values for Jarlskog
invariant and CP violating phase.}
\label{Observables0}
\end{table}

\bigskip Using the best-fit values given above, we get for NH and IH,
respectively, the following neutrino masses: 
\begin{equation}
m_{1}=0,\hspace{1cm}m_{2}\approx 8.72\mbox{meV},\hspace{1cm}m_{3}\approx
49.80\mbox{meV},\ \ \ \ \ \ \ \ \mbox{for NH}  \label{neutrinomassesNH}
\end{equation}%
\begin{equation}
m_{1}\approx 49.56\mbox{meV},\hspace{1cm}m_{2}\approx 48.79\mbox{meV},%
\hspace{1cm}m_{3}=0,\ \ \ \ \ \ \ \ \mbox{for IH}  \label{neutrinomassesIH}
\end{equation}%
The obtained and experimental values of the observables in the lepton sector
are shown in Table \ref{Observables0}. The experimental values of the
charged lepton masses, which are given at the $M_{Z}$ scale, have been taken
from Ref. \cite{Bora:2012tx} (which are similar to those in \cite%
{Xing:2007fb}), whereas the experimental values of the neutrino mass squared
splittings and leptonic mixing angles for both normal (NH) and inverted (IH)
mass hierarchies, are taken from Ref. \cite{Forero:2014bxa}. The obtained
charged lepton masses, neutrino mass squared splittings and lepton mixing
angles are in excellent agreement with the experimental data for both normal
and inverted neutrino mass hierarchies. We found a leptonic Dirac CP
violating phase close to $\frac{\pi }{2}$ and a Jarlskog invariant close to
about $3\times 10^{-2}$ for both normal and inverted neutrino mass hierarchy.

In order to study the sensitivity of the obtained values for the neutrino
mass squared splittings, under small variations around the best-fit values
(maximum variation of $+0.2$, minimum of $-0.2$), we show in Figures \ref{NH}
and \ref{IH} the correlations between $\sin^2\theta_{23}$ and $%
\sin^2\theta_{13}$, $\sin^2\theta_{12}$ and $\sin^2\theta_{13}$, $\Delta
m_{21}^{2}$ and $\Delta m_{31}^{2}$ for the case of normal and inverted
neutrino mass hierarchies, respectively. These Figures show that a slight
variation from the best-fit values, yields for several points of the
parameter space an important deviation in the values of the neutrino mass
squared splittings and leptonic mixing parameters, thus making it difficult
to reproduce their experimental values, especially for the case of inverted
neutrino mass hierarchy. Thus, the solution corresponding to the best fit
point is fine-tuned in the case of inverted neutrino mass hierarchy.
Addressing this problem requires to consider a discrete flavor group having
a triplet irreducible representation, such as, for example $A_4$, $S_4$ and $%
T^{\prime}$. This will yield more predictive textures for the lepton sector
thus solving the fine tuning problem. Addressing this issue requires a
carefull investigation that we left beyond the scope of the present paper
and is left for future studies.

Now we determine the effective Majorana neutrino mass parameter, which is
proportional to the neutrinoless double beta ($0\nu \beta \beta $) decay
amplitude. The effective Majorana neutrino mass parameter is given by: 
\begin{equation}
m_{\beta \beta }=\left\vert \sum_{j}U_{ek}^{2}m_{\nu _{k}}\right\vert ,
\label{mee}
\end{equation}%
where $U_{ej}^{2}$ and $m_{\nu _{k}}$ are the PMNS mixing matrix elements
and the Majorana neutrino masses, respectively.

We predict that the effective Majorana neutrino mass parameter for both
normal and inverted hierarchies: 
\begin{equation}
m_{\beta \beta }=\left\{ 
\begin{array}{l}
4\ \mbox{meV}\ \ \ \ \ \ \ \mbox{for \ \ \ \ NH} \\ 
48\ \mbox{meV}\ \ \ \ \ \ \ \mbox{for \ \ \ \ IH} \\ 
\end{array}%
\right.  \label{eff-mass-pred}
\end{equation}

Our obtained value $m_{\beta \beta }\approx 4\ \mbox{meV}$ for the effective
Majorana neutrino mass parameter in the case of normal hierarchy, is beyond
the reach of the present and forthcoming $0\nu \beta \beta $ decay
experiments. In the concerning to the inverted neutrino mass hierarchy, we
get the value $m_{\beta \beta }\approx 48$ for the Majorana neutrino mass
parameter, which is within the declared reach of the next-generation
bolometric CUORE experiment \cite{Alessandria:2011rc} or, more
realistically, of the next-to-next-generation ton-scale $0\nu \beta \beta $%
-decay experiments. The current best upper bound on the effective neutrino
mass is $m_{\beta \beta }\leq 160$ meV, which corresponds to $T_{1/2}^{0\nu
\beta \beta }(^{136}\mathrm{Xe})\geq 1.6\times 10^{25}$ yr at 90\% C.L, as
indicated by the EXO-200 experiment \cite{Auger:2012ar}. This bound will be
improved within a not too far future. The GERDA \textquotedblleft
phase-II\textquotedblright experiment \cite{Abt:2004yk,Ackermann:2012xja} 
is expected to reach 
\mbox{$T^{0\nu\beta\beta}_{1/2}(^{76}{\rm Ge}) \geq
2\times 10^{26}$ yr}, which corresponds to $m_{\beta \beta }\leq 100$ meV. A
bolometric CUORE experiment, using ${}^{130}Te$ \cite{Alessandria:2011rc},
is currently under construction and has an estimated sensitivity of about $%
T_{1/2}^{0\nu \beta \beta }(^{130}\mathrm{Te})\sim 10^{26}$ yr, which
corresponds to \mbox{$m_{\beta\beta}\leq 50$ meV.} Furthermore, there are
proposals for ton-scale next-to-next generation $0\nu \beta \beta $
experiments with $^{136}$Xe \cite{KamLANDZen:2012aa,Albert:2014fya} and $%
^{76}$Ge \cite{Abt:2004yk,Guiseppe:2011me} claiming sensitivities over $%
T_{1/2}^{0\nu \beta \beta }\sim 10^{27}$ yr, which corresponds to $m_{\beta
\beta }\sim 12-30$ meV. For a recent review, see for example Ref. \cite%
{Bilenky:2014uka}. Consequently, as follows from Eq. (\ref{eff-mass-pred}),
our model predicts $T_{1/2}^{0\nu \beta \beta }$ at the level of
sensitivities of the next generation or next-to-next generation $0\nu \beta
\beta $ experiments.


\section{Quark masses and mixing.}

\label{quarkmassesandmixing} From Eq. (\ref{YukawatermsQ}) and taking into
account that the VEV pattern of the $SU\left( 3\right) _{L}$ singlet scalar
fields is described by Eq. (\ref{VEV}), with the nonvanishing VEVs set to be
equal to $\lambda \Lambda $ (being $\Lambda $ the cutoff of our model)\ as
shown in Eq. (\ref{VEVsinglets}), it follows that the SM quark mass matrices
have the form: 
\begin{eqnarray}
M_{U} &=&\left( 
\begin{array}{ccc}
a_{1}^{\left( U\right) }\lambda ^{8} & 0 & 0 \\ 
0 & a_{2}^{\left( U\right) }\lambda ^{4} & a_{4}^{\left( U\right) }\lambda
^{2} \\ 
0 & 0 & a_{3}^{\left( U\right) }%
\end{array}%
\right) \frac{v}{\sqrt{2}},  \notag \\
M_{D} &=&\left( 
\begin{array}{ccc}
a_{1}^{\left( D\right) }\lambda ^{7} & a_{4}^{\left( D\right) }\lambda ^{6}
& a_{5}^{\left( D\right) }\lambda ^{6} \\ 
0 & a_{2}^{\left( D\right) }\lambda ^{5} & 0 \\ 
0 & 0 & a_{3}^{\left( D\right) }\lambda ^{3}%
\end{array}%
\right) \frac{v}{\sqrt{2}},  \label{Quarktextures}
\end{eqnarray}%
where $\lambda =0.225$ is one of the Wolfenstein parameters, $v=246$ GeV the
scale of electroweak symmetry breaking and $a_{i}^{\left( U\right) }$ ($%
i=1,2,3,4$) and $a_{j}^{\left( D\right) }$ ($j=1,2,3,4$) are $\mathcal{O}(1)$
parameters. From the SM quark mass textures given above, it follows that the
Cabbibo mixing as well as the mixing in the 1-3 plane emerges from the down
type quark sector, whereas the up type quark sector generates the quark
mixing angle in the 2-3 plane. Besides that, the low energy quark flavor
data indicates that the CP violating phase in the quark sector is associated
with the quark mixing angle in the 1-3 plane, as follows from the Standard
parametrization of the quark mixing matrix. Consequently, in order to get
quark mixing angles and a CP violating phase consistent with the
experimental data, we assume that all dimensionless parameters given in Eq. (%
\ref{Quarktextures}) are real, except for $a_{5}^{\left( D\right) }$, taken
to be complex.

Furthermore, the exotic quark masses read: 
\begin{equation}
m_{T}=y^{\left( T\right) }\frac{v_{\chi }}{\sqrt{2}},\hspace{1cm}%
m_{J^{1}}=y_{1}^{\left( J\right) }\frac{v_{\chi }}{\sqrt{2}}=\frac{%
y_{1}^{\left( J\right) }}{y^{\left( T\right) }}m_{T},\hspace{1cm}%
m_{J^{2}}=y_{2}^{\left( J\right) }\frac{v_{\chi }}{\sqrt{2}}=\frac{%
y_{2}^{\left( J\right) }}{y^{\left( T\right) }}m_{T}.  \label{mexotics}
\end{equation}

Since the charged fermion mass and quark mixing pattern arises from the
breaking of the $Z_{3}\otimes Z_{3}^{\prime }\otimes Z_{8}\otimes Z_{16}$
discrete group, and in order to simplify the analysis, we adopt a benchmark
where we set $a_{4}^{\left( D\right) }=a_{1}^{\left( D\right) }$ as well as $%
a_{1}^{\left( U\right) }=a_{3}^{\left( U\right) }=1$ and $a_{3}^{\left(
D\right) }=a_{2}^{\left( U\right) }$, motivated by naturalness arguments and
by the relation $m_{c}\sim m_{b}$, respectively. Then, we proceed to fit the
6 parameters $a_{2}^{\left( U\right) }$, $a_{4}^{\left( U\right) }$, $%
a_{1}^{\left( D\right) }$, $a_{2}^{\left( D\right) }$, $a_{5}^{\left(
D\right) }$ and the phase $\tau $, to reproduce the 10 physical observables
of the quark sector{, i.e., the six quark masses, the three mixing angles
and the CP violating phase. The obtained values for the quark masses, the
three quark mixing angles and the CP violating phase $\delta $ in Table \ref%
{Tab} correspond to the best fit values: }

\begin{eqnarray}
a_{2}^{\left( U\right) } &\simeq &1.43,\hspace{1cm}a_{4}^{\left( U\right)
}\simeq 0.80,\hspace{1cm}a_{1}^{\left( D\right) }\simeq 0.58,  \notag \\
a_{2}^{\left( D\right) } &\simeq &0.57,\hspace{1cm}\left\vert a_{5}^{\left(
D\right) }\right\vert \simeq 0.44,\hspace{1cm}\tau =68^{\circ }.
\end{eqnarray}

\begin{table}[tbh]
\begin{center}
\begin{tabular}{c|l|l}
\hline\hline
Observable & Model value & Experimental value \\ \hline
$m_{u}(MeV)$ & \quad $1.16$ & \quad $1.45_{-0.45}^{+0.56}$ \\ \hline
$m_{c}(MeV)$ & \quad $641$ & \quad $635\pm 86$ \\ \hline
$m_{t}(GeV)$ & \quad $174$ & \quad $172.1\pm 0.6\pm 0.9$ \\ \hline
$m_{d}(MeV)$ & \quad $2.9$ & \quad $2.9_{-0.4}^{+0.5}$ \\ \hline
$m_{s}(MeV)$ & \quad $59.2$ & \quad $57.7_{-15.7}^{+16.8}$ \\ \hline
$m_{b}(GeV)$ & \quad $2.85$ & \quad $2.82_{-0.04}^{+0.09}$ \\ \hline
$\sin \theta _{12}$ & \quad $0.225$ & \quad $0.225$ \\ \hline
$\sin \theta _{23}$ & \quad $0.0407$ & \quad $0.0412$ \\ \hline
$\sin \theta _{13}$ & \quad $0.00352$ & \quad $0.00351$ \\ \hline
$\delta $ & \quad $68^{\circ }$ & \quad $68^{\circ }$ \\ \hline\hline
\end{tabular}%
\end{center}
\caption{Model and experimental values of the quark masses and CKM
parameters.}
\label{Tab}
\end{table}

The obtained quark masses, quark mixing angles and CP violating phase are
consistent with the experimental data. Let us note, that despite the
aforementioned simplifying assumptions that allow us to eliminate some of
the free parameters, a good fit with the low energy quark flavor data is
obtained, showing that our model is indeed capable of a very good fit to the
experimental data of the physical observables for the quark sector. The
obtained and experimental values for the physical observables of the quark
sector are reported in Table \ref{Tab}. We use the experimental values of
the quark masses at the $M_{Z}$ scale, from Ref. \cite{Bora:2012tx} (which
are similar to those in \cite{Xing:2007fb}), whereas the experimental values
of the CKM parameters are taken from Ref. \cite{Agashe:2014kda}. We have
numerically checked that a slight deviation from the best-fit values, keeps
all the obtained SM quark masses, with the exception of the bottom quark
mass, inside the $3\sigma$ experimentally allowed range. We checked that
small variations around the best fit values, keep most of the resulting
values of the bottom quark mass inside the $3\sigma$ experimentally allowed
range. The values outside the $3\sigma$ experimentally allowed range are
close to the lower and upper experimental bounds of the bottom quark mass.
Consequently, our model is very predictive for the quark sector.

On the other hand, from the SM quark textures, it follows that in order to
obtain realistic SM quark masses and mixing angles without requiring a
strong hierarchy among the Yukawa couplings, one should have $v_{\rho }\sim
v_{\eta }$, which implies that $\tan \beta \sim \mathcal{O}(1)$.
Furthermore, as the $h^{0}b\bar{b}$ coupling is proportional to $\frac{\sin
\alpha }{\cos \beta }$, in order to get a $h^{0}b\bar{b}$ coupling close to
the SM expectation, we have $\alpha \sim \beta \pm \frac{\pi }{2}$. In what
follows we briefly comment about the phenomenological implications of our
model in the concerning to the flavor changing processes involving quarks.
As previously mentioned, the different $Z_{3}$ charge assignments for SM and
exotic right handed quark fields imply the absence of mixing between them.
The absence of mixings between the SM and exotic quarks will imply that the
exotic fermions will not exhibit flavor changing decays into SM quarks and
gauge (or Higgs) bosons. After being pair produced they will decay into the
SM quarks and the intermediate states of heavy gauge bosons, which in turn
decay into the pairs of the SM fermions, see e.g. \cite{Cabarcas:2008ys}.
The precise signature of the decays of the exotic quarks depends on details
of the spectrum and other parameters of the model. The present lower bounds
from the LHC on the masses of the $Z^{\prime }$ gauge bosons in the $3\text{-%
}3\text{-}1$ models reach around $2.5$ \text{TeV} \cite{Salazar:2015gxa}.
One can translate these bounds on the order of magnitude of the scale $%
v_{\chi }$ of breaking of the $SU(3)_{C}\otimes SU\left( 3\right)
_{L}\otimes U\left( 1\right) _{X}\otimes Z_{3}$ symmetry. These exotic
quarks can be produced at the LHC via Drell-Yan proccesses mediated by
charged gauge bosons, where the final states will include the exotic $T$
quark with a SM down type quark as well as any of the exotic $J^{1}$ or $%
J^{2}$ quarks with a SM up type quark. It would be interesting to perform a
detailed study of the exotic quark production at the LHC, the exotic quark
decay modes and the flavor changing top quark decays. This is beyond the
scope of this work and is left for future studies.

\section{Conclusions}

\label{conclusions}

We have constructed an extension of the 3-3-1 model with $\beta =-\frac{1}{%
\sqrt{3}}$, based on the extended $SU(3)_{C}\otimes SU(3)_{L}\otimes
U(1)_{X}\otimes S_{3}\otimes Z_{3}\otimes Z_{3}^{\prime }\otimes
Z_{8}\otimes Z_{16}$ symmetry. Our $S_{3}$ flavor 3-3-1 model, which at low
energies reduces to the 3-3-1 model with right handed neutrinos, where $%
\beta =-\frac{1}{\sqrt{3}}$, is in agreement with the current data on SM
fermion masses and mixing. The $S_{3}$, $Z_{3}$, $Z_{3}^{\prime }$ and $%
Z_{8} $ discrete groups reduce the number of the model parameters.
Specifically, the $Z_{3}$ and $Z_{8}$ symmetries determines the allowed
entries of the charged lepton mass matrix. Furthermore, the $Z_{3}$ symmetry
decouples the SM quarks from the exotic quarks. The $Z_{3}^{\prime }$
symmetry selects the allowed entries of the SM quark mass matrices. The $%
Z_{16}$ symmetry generates the hierarchy among charged fermion masses and
quark mixing angles that yields the observed charged fermion mass and quark
mixing pattern. We assumed that the $SU(3)_{L}$ scalar singlets having
Yukawa interactions with the right handed Majorana neutrinos acquire VEVs
much smaller than the electroweak symmetry breaking scale, then providing
very small masses to these Majorana neutrinos, and thus giving rise to an
inverse seesaw mechanism of active neutrino masses. 
The smallness of the active neutrino masses is attributed to their scaling
with inverse powers of the high energy cutoff $\Lambda $ as well as well as
by their linear dependence on the very small VEVs of the $SU(3)_{L}$
singlets $\varphi _{j}$ ($j=1,2$), assumed to be of the same order of
magnitude. We found for these VEVs, the estimate $v_{\varphi }\sim 0.5$ GeV.
The observed hierarchy of SM charged fermion masses and quark mixing matrix
elements arises from the breaking of the $Z_{3}\otimes Z_{3}^{\prime
}\otimes Z_{8}\otimes Z_{16}$ discrete group at very high energy.
Furthermore, the model features a leptonic Dirac CP violating phase close to 
$\frac{\pi }{2}$ and a Jarlskog invariant close to about $3\times 10^{-2}$
for both normal and inverted neutrino mass hierarchy. In addition, under the
assumption that the exotic $T$, $J^{1}$ and $J^{2}$ quarks are lighter than
the $H_{2}^{0}$ and $\overline{H}_{2}^{0}$\ neutral Higgs bosons, our model
predicts the absence of the flavor changing neutral exotic quark decays,
which implies that they can be searched at the LHC via their decay into the
SM quarks and the intermediate states of heavy gauge bosons, which in turn
decay into the pairs of the SM fermions, see e.g. \cite{Cabarcas:2008ys}.
Possible directions for future work along these lines would be to study the
constraints on the heavy charged gauge boson masses in our model arising
from the upper bound on the branching fraction for the flavor changing top
quark decays, the oblique parameters, the $Zb\overline{b} $ vertex and the
Higgs diphoton signal strength. The heavy exotic quark production at the LHC
may be useful to study. Finally we briefly comment about a posible large
discrete symmetry group that could be used to embed the $S_{3}\otimes
Z_{3}\otimes Z_{3}^{\prime }\otimes Z_{8}\otimes Z_{16}$ discrete symmetry
of our model. Considering that the discrete group $\Delta \left(
6N^{2}\right) $ is isomorphic to$\ (Z_{N}\times Z_{N}^{\prime })\rtimes S_{3}
$ \cite{Ishimori:2010au} and the fact the $Z_{24}$ discrete group is the
smallest cyclic group that contains both the $Z_{3}$ and $Z_{8}$ symmetries,
it follows that the $S_{3}\otimes Z_{3}\otimes Z_{3}^{\prime }\otimes
Z_{8}\otimes Z_{16}$ discrete group of our model can be embedded in the $%
\Delta \left( 6N^{2}\right) =\Delta $ $\left( 3456\right) $ discrete group
(where $N=24$). It would be interesting to implement the $\Delta \left(
6N^{2}\right) $ discrete symmetry in the 331 model and to study its
implications on fermion masses and mixings. All these studies require
carefull investigations that we left outside the scope of this work.

\section*{Acknowledgments}

A.E.C.H was supported by Fondecyt (Chile), Grant No. 11130115, by DGIP
internal Grant No. 111458 and by Proyecto Basal FB0821. R. M. and F.O. were supported by El Patrimonio
Aut\'onomo Fondo Nacional de Financiamiento para la Ciencia, la Tecnolog%
\'{\i}a y la Innovaci\'on Fransisco Jos\'e de Caldas programme of
COLCIENCIAS in Colombia. The visit of R.M to Universidad T\'{e}cnica
Federico Santa Mar\'{\i}a was supported by Fondecyt (Chile), Grant No.
11130115. We are very grateful to an anonymous referee for the careful reading of this article and for very valuable comments and suggestions. 

\appendix

\section{The product rules for $S_{3}$.}

\label{S3}

The $S_{3}$ group has three irreducible representations: $\mathbf{1}$, $%
\mathbf{1}^{\prime }$ and $\mathbf{2}$. Denoting the basis vectors for two $%
S_{3}$ doublets as $\left( x_{1},x_{2}\right) ^{T}$\ and $\left(
y_{1},y_{2}\right) ^{T}$ and $y^{\prime }$ a non trivial $S_{3}$ singlet,
the $S_{3}$ multiplication rules are \cite{Ishimori:2010au}:

\begin{equation}
\left( 
\begin{array}{c}
x_{1} \\ 
x_{2}%
\end{array}%
\right) _{\mathbf{2}}\otimes \left( 
\begin{array}{c}
y_{1} \\ 
y_{2}%
\end{array}%
\right) _{\mathbf{2}}=\left( x_{1}y_{1}+x_{2}y_{2}\right) _{\mathbf{1}%
}+\left( x_{1}y_{2}-x_{2}y_{1}\right) _{\mathbf{1}^{\prime }}+\left( 
\begin{array}{c}
x_{2}y_{2}-x_{1}y_{1} \\ 
x_{1}y_{2}+x_{2}y_{1}%
\end{array}%
\right) _{\mathbf{2}},  \label{6}
\end{equation}%
\begin{equation}
\left( 
\begin{array}{c}
x_{1} \\ 
x_{2}%
\end{array}%
\right) _{\mathbf{2}}\otimes \left( y^{\prime }\right) _{\mathbf{1}^{\prime
}}=\left( 
\begin{array}{c}
-x_{2}y^{\prime } \\ 
x_{1}y^{\prime }%
\end{array}%
\right) _{\mathbf{2}},\hspace{1cm}\hspace{1cm}\left( x^{\prime }\right) _{%
\mathbf{1}^{\prime }}\otimes \left( y^{\prime }\right) _{\mathbf{1}^{\prime
}}=\left( x^{\prime }y^{\prime }\right) _{\mathbf{1}}.  \label{7}
\end{equation}

With these multiplication rules we have to assign to the scalar fields in
the $S_{3}$ irreps and build the corresponding scalar potential invariant
under the symmetry group.

\section{Stability conditions of the low energy scalar potential.}

\label{stability} In what follows we are going to determine the conditions required to have
a stable scalar potential by following the method described in Ref. \cite{Maniatis:2006fs}. The gauge invariant and renormalizable low energy scalar
potential as a function of the fields $\phi _{1}=\chi $, $\phi _{2}=\rho $
and $\phi _{3}=\eta $ is a linear hermitian combination of the following
terms: 
\begin{equation}
\phi _{i}\phi _{j},\;\;\;\;\;\;\phi _{i}\phi _{j}\phi _{k}\phi _{l}
\end{equation}%
%
%
%
%
%
%
%
%
%
%
%
%
%
%
%
%
%
%
%
%
%
%
%
%
%
%
%
%
%
%
%
where $i,j,k,l=\phi _{1}$, $\phi _{2}$ and $\phi _{3}$. To discuss the
stability of the potential, its minimum, and its gauge invariance one can
make the following arrangement of the scalar fields by using $2\times 2$
hermitian matrices as follows: 
\begin{eqnarray}
\widetilde{K}_{(\phi _{i}\phi _{j})} &=&\left( 
\begin{array}{cc}
\phi _{i}^{\dagger }\phi _{i} & \phi _{i}^{\dagger }\phi _{j} \\ 
\phi _{j}^{\dagger }\phi _{i} & \phi _{j}^{\dagger }\phi _{j}%
\end{array}%
\right) ,  \notag \\
&=&\frac{1}{2}\left( K_{0\left( \phi _{i}\phi _{j}\right) }1_{2\times
2}+K_{a\left( \phi _{i}\phi _{j}\right) }\sigma ^{a}\right)
\end{eqnarray}%
%
%
%
%
%
where $(\phi _{i}\phi _{j})=\rho \eta ,\rho \chi ,\eta \chi $, $\sigma ^{a}$
($a=1,2,3$) are the Pauli matrices and $1_{2\times 2}$ is the identity
matrix. From the previous expressions one can build the following bilinear
terms as functions of the scalar fields: 
\begin{eqnarray}
K_{0\left( \phi _{i}\phi _{j}\right) } &=&\phi _{i}^{\dagger }\phi _{i}+\phi
_{j}^{\dagger }\phi _{j},\hspace{1cm}\hspace{1cm}  \notag \\
K_{a\left( \phi _{i}\phi _{j}\right) } &=&\sum_{i,j}\left( \phi
_{i}^{\dagger }\phi _{j}\right) \sigma _{ij}^{a}.
\end{eqnarray}%
\quad The properties of the potential can be analyzed in terms of $%
K_{0\left( \phi _{i}\phi _{j}\right) }$ and $\vec{K}_{\left( \phi _{i}\phi
_{j}\right) }$ with $\phi _{i}\phi _{j}=\rho \eta ,\rho \chi ,\eta \chi $ in
the domain $K_{0}\geq 0$ y $K_{0}^{2}\geq \vec{K}^{2}$. Defining $\vec{\kappa%
}=\vec{K}/K_{0}$ the potential can be written as
\begin{eqnarray}
V &=&V_{2}+V_{4},  \notag \\
V_{2} &=&\sum_{(\phi _{i}\phi _{j})}K_{0(\phi _{i}\phi _{j})}\vec{J}_{2(\phi
_{i}\phi _{j})}(\vec{\kappa}),\;\;\;\;  \notag \\
\vec{J}_{2{(\phi _{i}\phi _{j})}}(\vec{\kappa}) &=&\xi _{0(\phi _{i}\phi
_{j})}+\vec{\xi}_{(\phi _{i}\phi _{j})}^{T}\vec{\kappa}_{(\phi _{i}\phi
_{j})},  \notag \\
V_{4} &=&\sum_{(\phi _{i}\phi _{j})}K_{0(\phi _{i}\phi _{j})}^{2}\vec{J}%
_{4(\phi _{i}\phi _{j})}(\vec{\kappa}),\;\;\;  \label{potential} \\
\;\vec{J}_{4{(\phi _{i}\phi _{j})}}(\vec{\kappa}) &=&\eta _{00(\phi _{i}\phi
_{j})}+2\vec{\eta}_{(\phi _{i}\phi _{j})}^{T}\vec{\kappa}_{(\phi _{i}\phi
_{j})}  \notag \\
&&+\vec{\kappa}_{(\phi _{i}\phi _{j})}^{T}E_{(\phi _{i}\phi _{j})}\vec{\kappa%
}_{(\phi _{i}\phi _{j})},  \notag
\end{eqnarray}%
where $E_{(\phi _{i}\phi _{j})}$ is a $3\times 3$ matrix and the functions $J_{2{(\phi _{i}\phi _{j})}}(\vec{\kappa})$ and $J_{4{(\phi _{i}\phi _{j})}}(%
\vec{\kappa})$ are defined in the domain $|\vec{\kappa}|\leq 1$. The
stability of the scalar potential requires that it has to be bounded from
below. The stability is determined from the behavior of $V$ in the limit $%
K_{0}\rightarrow \infty $, i.e., 
\begin{equation}
J_{4{(\phi _{i}\phi _{j})}}(\vec{\kappa})\geq 0,
\end{equation}%
for all $|\vec{\kappa}|\leq 1$. To impose $J_{4{(\phi _{i}\phi _{j})}}(\vec{%
\kappa})$ to be positively defined it is enough to consider the values of
all stationary points in the domain $|\kappa |<1$ and $|\kappa |=1$. This
results in a bound for $\eta _{00{(\phi _{i}\phi _{j})}}$, $\vec{\eta}_{0{%
(\phi _{i}\phi _{j})}}$ and $E{(\phi _{i}\phi _{j})}$, which parametrize the
quartic terms of the potential included in $V_{4}$.

For $|\vec{\kappa}|<1$ the stationary points should satisfy 
\begin{equation}
E\vec{\kappa}_{(\phi _{i}\phi _{j})}=-\vec{\eta}_{(\phi _{i}\phi
_{j})},\;\;\;\;\;\;|\vec{\kappa}|<1.
\end{equation}

For the case where $\det E\neq 0$, the following relation is obtained: 
\begin{equation}
\;J_{4(\phi _{i}\phi _{j})}(\vec{\kappa})|_{est}=\eta _{00{(\phi _{i}\phi
_{j})}}-\vec{\eta}_{(\phi _{i}\phi _{j})}^{T}E^{-1}\vec{\eta}_{(\phi
_{i}\phi _{j})}.
\end{equation}

For $|\vec{\kappa}|=1$ the stationary points are obtained from the function: 
\begin{equation}
F_{4(\phi _{i}\phi _{j})}(\vec{\kappa})=J_{4(\phi _{i}\phi _{j})}(\kappa
)+u(1-\vec{\kappa}^{2}),
\end{equation}%
where $u$ is a Lagrange multiplier that satisfies the following condition 
\begin{eqnarray}
(E_{(\phi _{i}\phi _{j})}-u)\vec{\kappa} &=&-\vec{\eta}_{(\phi _{i}\phi
_{j})},  \notag \\
\;J_{4(\phi _{i}\phi _{j})}(\vec{\kappa})|_{est} &=&u+\eta _{00{(\phi
_{i}\phi _{j})}} \\
&&-\vec{\eta}_{(\phi _{i}\phi _{j})}^{T}(E_{(\phi _{i}\phi _{j})}-u)^{-1}%
\vec{\eta}_{(\phi _{i}\phi _{j})}.  \notag
\end{eqnarray}%
The stationary points of $J_{4(\phi _{i}\phi _{j})}(\kappa )$ for $|\kappa
|\leq 1$ can be obtained from: 
\begin{eqnarray}
f_{(\phi _{i}\phi _{j})}(u) &=&J_{4{(\phi _{i}\phi _{j})}}(\vec{\kappa}%
)|_{est}>0,  \notag \\
f_{(\phi _{i}\phi _{j})}^{\prime }\left( u\right) &>&0.
\label{stabilityconditions}
\end{eqnarray}%
Considering that the quartic terms of the scalar potential are dominant
when the vacuum expectation values of the scalar fields take large values,
these terms will be the most relevant to analyze the stability of the scalar
potential. Following the method described in Ref. \cite{Maniatis:2006fs}, we
proceed to rewrite the quartic terms of the scalar potential in terms of
bilinear combinations of the scalar fields. To this end, the bilinear
combinations of the scalar fields are included in the following matrices: 
\begin{eqnarray}
\widetilde{K}_{\rho \eta } &=&\left( 
\begin{array}{cc}
\rho ^{\dagger }\rho & \eta ^{\dagger }\rho \\ 
\rho ^{\dagger }\eta & \eta ^{\dagger }\eta%
\end{array}%
\right) =\frac{1}{2}\left( K_{0\left( \rho \eta \right) }1_{2\times
2}+K_{a\left( \rho \eta \right) }\sigma ^{a}\right) ,  \notag \\
\widetilde{K}_{\rho \chi } &=&\left( 
\begin{array}{cc}
\rho ^{\dagger }\rho & \chi ^{\dagger }\rho \\ 
\rho ^{\dagger }\chi & \chi ^{\dagger }\chi%
\end{array}%
\right) =\frac{1}{2}\left( K_{0\left( \rho \chi \right) }1_{2\times
2}+K_{a\left( \rho \chi \right) }\sigma ^{a}\right) ,  \notag \\
\widetilde{K}_{\eta \chi } &=&\left( 
\begin{array}{cc}
\eta ^{\dagger }\eta & \chi ^{\dagger }\eta \\ 
\eta ^{\dagger }\chi & \chi ^{\dagger }\chi%
\end{array}%
\right) =\frac{1}{2}\left( K_{0\left( \eta \chi \right) }1_{2\times
2}+K_{a\left( \eta \chi \right) }\sigma ^{a}\right) ,  \notag \\
&&  \label{S1}
\end{eqnarray}%
where $\sigma ^{a}$ ($a=1,2,3$) are the Pauli matrices and $1_{2\times 2}$
is the $2\times 2$ identity matrix. From the previous expressions, we find
that the bilinear combinations of the scalar fields appearing in Eq. (\ref%
{S1}) are given by: 
\begin{eqnarray}
K_{0\left( \rho \eta \right) } &=&\rho ^{\dagger }\rho +\eta ^{\dagger }\eta
,\hspace{0.5cm}K_{0\left( \rho \chi \right) }=\rho ^{\dagger }\rho +\chi
^{\dagger }\chi ,\hspace{0.5cm}  \notag \\
K_{0\left( \eta \chi \right) } &=&\eta ^{\dagger }\eta +\chi ^{\dagger }\chi
,  \label{S2} \\
K_{a\left( \rho \eta \right) } &=&\left( \rho ^{\dagger }\rho \right) \sigma
_{11}^{a}+\left( \eta ^{\dagger }\eta \right) \sigma _{22}^{a}+\left( \rho
^{\dagger }\eta \right) \sigma _{12}^{a}+\left( \eta ^{\dagger }\rho \right)
\sigma _{21}^{a},  \notag \\
K_{a\left( \rho \chi \right) } &=&\left( \rho ^{\dagger }\rho \right) \sigma
_{11}^{a}+\left( \chi ^{\dagger }\chi \right) \sigma _{22}^{a}+\left( \rho
^{\dagger }\chi \right) \sigma _{12}^{a}+\left( \chi ^{\dagger }\rho \right)
\sigma _{21}^{a},  \notag \\
K_{a\left( \eta \chi \right) } &=&\left( \eta ^{\dagger }\eta \right) \sigma
_{11}^{a}+\left( \chi ^{\dagger }\chi \right) \sigma _{22}^{a}+\left( \eta
^{\dagger }\chi \right) \sigma _{12}^{a}+\left( \chi ^{\dagger }\eta \right)
\sigma _{21}^{a}.  \notag
\end{eqnarray}

Since the stability of the scalar potential is determined from its quartic
terms, the stationary solutions consistent with a stable scalar potential
are described by the following functions: 
\begin{eqnarray}
f_{\rho \eta }\left( u\right) &=&u+E_{00\left( \rho \eta \right)
}-E_{a\left( \rho \eta \right) }\left( E_{\left( \rho \eta \right)
}-u1_{3\times 3}\right) _{ab}^{-1}E_{b\left( \rho \eta \right) },  \notag \\
f_{\rho \chi }\left( u\right) &=&u+E_{00\left( \rho \chi \right)
}-E_{a\left( \rho \chi \right) }\left( E_{\left( \rho \chi \right)
}-u1_{3\times 3}\right) _{ab}^{-1}E_{b\left( \rho \chi \right) },  \notag \\
f_{\eta \chi }\left( u\right) &=&u+E_{00\left( \eta \chi \right)
}-E_{a\left( \eta \chi \right) }\left( E_{\left( \eta \chi \right)
}-u1_{3\times 3}\right) _{ab}^{-1}E_{b\left( \eta \chi \right) },  \notag \\
&&  \label{S5}
\end{eqnarray}%
where, for the $\rho $ and $\eta $ fields, we have 
\begin{eqnarray}
E_{00\left( \rho \eta \right) } &=&\frac{\lambda _{2}+\lambda
_{3}+\lambda _{6}}{4},\hspace{1cm}  \notag \\
E_{a\left( \rho \eta \right) } &=&\frac{\lambda _{2}-\lambda
_{3}}{4}\delta _{a3},  \notag \\
E_{\left( \rho \eta \right) } &=&\frac{1}{4}\left( 
\begin{array}{ccc}
\lambda _{9} & 0 & 0 \\ 
0 & \lambda _{9} & 0 \\ 
0 & 0 & \lambda _{2}+\lambda _{3}-\lambda _{6}%
\end{array}%
\right) ,  \label{MS1}
\end{eqnarray}

In the same manner, for the multiplets $\rho $ and $\chi $, the expressions
are 
\begin{eqnarray}
E_{00\left( \rho \chi \right) } &=&\frac{\lambda _{1}+\lambda _{2}+\lambda
_{4}}{4},\hspace{1cm}  \notag \\
E_{a\left( \rho \chi \right) } &=&\frac{\lambda _{1}-\lambda _{2}}{4}\delta _{a3},  \notag \\
E_{\left( \rho \chi \right) } &=&\frac{1}{4}\left( 
\begin{array}{ccc}
\lambda _{8} & 0 & 0 \\ 
0 & \lambda _{8} & 0 \\ 
0 & 0 & \lambda _{1}+\lambda _{2}-\lambda _{4}%
\end{array}%
\right) .  \label{MS2}
\end{eqnarray}%
Similarly, for the $\eta $ and $\chi $ fields, we find: 
\begin{equation}
E_{00\left( \eta \chi \right) }=\frac{\lambda _{1}+\lambda _{3}+\lambda
_{5}}{4},\hspace{0.5cm}E_{a\left( \eta
\chi \right) }=\frac{\lambda _{1}-\lambda _{3}}{4}\delta _{a3},\hspace{0.5cm}E_{\left( \eta \chi \right) }=\frac{1}{4}\left( 
\begin{array}{ccc}
\lambda_{7} & 0 & 0 \\ 
0 & \lambda_{7} & 0 \\ 
0 & 0 & \lambda_{1}+\lambda_{3}-\lambda_{5}%
\end{array}%
\right) .  \label{MS3}
\end{equation}

Following Ref. \cite{Maniatis:2006fs}, we determine the stability of the
scalar potential from the conditions: 
\begin{equation}
f_{\rho \eta }\left( u\right) >0,\hspace{0.5cm}f_{\rho \chi }\left( u\right)
>0,\hspace{0.5cm}f_{\eta \chi }\left( u\right) >0.  \label{S6}
\end{equation}

We use the theorem of stability of the scalar potential of Ref. \cite{Maniatis:2006fs} to determine the stability conditions of the scalar
potential. To this end, the condition $f_{\rho \eta }\left( u\right) >0$ is
analyzed for the set of values of $u$ which include the $0$, (since $f%
{\acute{}}%
_{\rho \eta }\left( 0\right) >0$) the roots $u_{\rho \eta }^{\left( 1\right)
}$ and $u_{\rho \eta }^{\left( 2\right) }$ of the equation $f%
{\acute{}}%
_{\rho \eta }\left( u\right) =0$ and the eigenvalues $\widetilde{E}_{\left(
\rho \eta \right) }^{\left( a\right) }$\ of the matrix $E_{\left( \rho \eta
\right) }$ where $f_{\rho \eta }\left( \widetilde{E}_{\left( \rho \eta
\right) }^{\left( a\right) }\right) $ is finite and $f%
{\acute{}}%
_{\rho \eta }\left( \widetilde{E}_{\left( \rho \eta \right) }^{\left(
a\right) }\right) \geq 0$ . We proceed in a similar way when analyzing the
conditions $f_{\rho \chi }\left( u\right) >0$ and $f_{\eta \chi }\left(
u\right) >0$.

Therefore, the scalar potential is stable when the following conditions are
fulfilled: 
\begin{eqnarray}
\lambda _{1} &>&0,\hspace{1cm}\lambda _{2}>0,\hspace{1cm}\lambda _{3}>0,\notag \\
\lambda _{4}+\lambda _{8} &\gtrless&2\sqrt{\lambda_{1}\lambda_{2}},\hspace{1cm}\lambda _{4}+\lambda _{8}\gtrless\lambda _{1}+\lambda _{2},\hspace{1cm}2\sqrt{\lambda_{1}\lambda_{2}}\gtrless\lambda _{4},\hspace{1cm}\lambda _{1}+\lambda _{2}\gtrless\lambda _{4},\notag \\
\lambda _{5}+\lambda _{7} &\gtrless&2\sqrt{\lambda_{1}\lambda_{3}},\hspace{1cm}\lambda _{5}+\lambda _{7}\gtrless\lambda _{1}+\lambda _{3},\hspace{1cm}2\sqrt{\lambda_{1}\lambda_{3}}\gtrless\lambda_{5},\hspace{1cm}\lambda _{1}+\lambda_{3}\gtrless\lambda _{5},\notag \\
\lambda _{6}+\lambda _{9} &\gtrless&2\sqrt{\lambda_{2}\lambda_{3}},\hspace{1cm}\lambda _{6}+\lambda _{9}\gtrless\lambda _{2}+\lambda _{3},\hspace{1cm}2\sqrt{\lambda_{2}\lambda_{3}}\gtrless\lambda _{6},\hspace{1cm}\lambda _{2}+\lambda _{3}\gtrless\lambda_{6}.\label{S7a}
\end{eqnarray}%
%
%

Furthermore, having masses $m_{H_{1}^{\pm }}^{2}$, $m_{H_{1}^{0}}^{2}$ and $m_{H_{3}^{0}}^{2}$ positively defined requires the following condition: 
\begin{equation}
f>0.
\end{equation}




\begin{thebibliography}{9}
\bibitem{Aad:2012tfa} 
  G.~Aad {\it et al.} [ATLAS Collaboration],
  Phys.\ Lett.\ B {\bf 716}, 1 (2012)
  doi:10.1016/j.physletb.2012.08.020
  [arXiv:1207.7214 [hep-ex]].



\bibitem{Chatrchyan:2012xdj} 
  S.~Chatrchyan {\it et al.} [CMS Collaboration],
  Phys.\ Lett.\ B {\bf 716}, 30 (2012)
  doi:10.1016/j.physletb.2012.08.021
  [arXiv:1207.7235 [hep-ex]].



\bibitem{An:2012eh} 
  F.~P.~An {\it et al.} [Daya Bay Collaboration],
  Phys.\ Rev.\ Lett.\  {\bf 108}, 171803 (2012)
  doi:10.1103/PhysRevLett.108.171803
  [arXiv:1203.1669 [hep-ex]].



\bibitem{Abe:2011sj} 
  K.~Abe {\it et al.} [T2K Collaboration],
  Phys.\ Rev.\ Lett.\  {\bf 107}, 041801 (2011)
  doi:10.1103/PhysRevLett.107.041801
  [arXiv:1106.2822 [hep-ex]].



\bibitem{Adamson:2011qu} 
  P.~Adamson {\it et al.} [MINOS Collaboration],
  Phys.\ Rev.\ Lett.\  {\bf 107}, 181802 (2011)
  doi:10.1103/PhysRevLett.107.181802
  [arXiv:1108.0015 [hep-ex]].



\bibitem{Abe:2011fz} 
  Y.~Abe {\it et al.} [Double Chooz Collaboration],
  Phys.\ Rev.\ Lett.\  {\bf 108}, 131801 (2012)
  doi:10.1103/PhysRevLett.108.131801
  [arXiv:1112.6353 [hep-ex]].



\bibitem{Ahn:2012nd} 
  J.~K.~Ahn {\it et al.} [RENO Collaboration],
  Phys.\ Rev.\ Lett.\  {\bf 108}, 191802 (2012)
  doi:10.1103/PhysRevLett.108.191802
  [arXiv:1204.0626 [hep-ex]].



\bibitem{Forero:2014bxa} 
  D.~V.~Forero, M.~Tortola and J.~W.~F.~Valle,
  Phys.\ Rev.\ D {\bf 90}, no. 9, 093006 (2014)
  doi:10.1103/PhysRevD.90.093006
  [arXiv:1405.7540 [hep-ph]].



\bibitem{Agashe:2014kda} 
  K.~A.~Olive {\it et al.} [Particle Data Group Collaboration],
  Chin.\ Phys.\ C {\bf 38}, 090001 (2014).
  doi:10.1088/1674-1137/38/9/090001



\bibitem{Fritzsch:1977za} 
  H.~Fritzsch,
  Phys.\ Lett.\ B {\bf 70}, 436 (1977).
  doi:10.1016/0370-2693(77)90408-7



\bibitem{Fukuyama:1997ky} 
  T.~Fukuyama and H.~Nishiura,
  hep-ph/9702253.



\bibitem{Du:1992iy} 
  D.~s.~Du and Z.~z.~Xing,
  Phys.\ Rev.\ D {\bf 48}, 2349 (1993).
  doi:10.1103/PhysRevD.48.2349



\bibitem{Barbieri:1994kw} 
  R.~Barbieri, G.~R.~Dvali, A.~Strumia, Z.~Berezhiani and L.~J.~Hall,
  Nucl.\ Phys.\ B {\bf 432}, 49 (1994)
  doi:10.1016/0550-3213(94)90593-2
  [hep-ph/9405428].



\bibitem{Peccei:1995fg} 
  R.~D.~Peccei and K.~Wang,
  Phys.\ Rev.\ D {\bf 53}, 2712 (1996)
  doi:10.1103/PhysRevD.53.2712
  [hep-ph/9509242].



\bibitem{Fritzsch:1999ee} 
  H.~Fritzsch and Z.~z.~Xing,
  Prog.\ Part.\ Nucl.\ Phys.\  {\bf 45}, 1 (2000)
  doi:10.1016/S0146-6410(00)00102-2
  [hep-ph/9912358].



\bibitem{Roberts:2001zy} 
  R.~G.~Roberts, A.~Romanino, G.~G.~Ross and L.~Velasco-Sevilla,
  Nucl.\ Phys.\ B {\bf 615}, 358 (2001)
  doi:10.1016/S0550-3213(01)00408-4
  [hep-ph/0104088].



\bibitem{Nishiura:2002ei} 
  H.~Nishiura, K.~Matsuda, T.~Kikuchi and T.~Fukuyama,
  Phys.\ Rev.\ D {\bf 65}, 097301 (2002)
  doi:10.1103/PhysRevD.65.097301
  [hep-ph/0202189].



\bibitem{deMedeirosVarzielas:2005ax} 
  I.~de Medeiros Varzielas and G.~G.~Ross,
  Nucl.\ Phys.\ B {\bf 733}, 31 (2006)
  doi:10.1016/j.nuclphysb.2005.10.039
  [hep-ph/0507176].



\bibitem{Carcamo:2006dp} 
  A.~E.~Carcamo Hernandez, R.~Martinez and J.~A.~Rodriguez,
  Eur.\ Phys.\ J.\ C {\bf 50}, 935 (2007)
  doi:10.1140/epjc/s10052-007-0264-0
  [hep-ph/0606190].



\bibitem{Kajiyama:2007gx} 
  Y.~Kajiyama, M.~Raidal and A.~Strumia,
  Phys.\ Rev.\ D {\bf 76}, 117301 (2007)
  doi:10.1103/PhysRevD.76.117301
  [arXiv:0705.4559 [hep-ph]].



\bibitem{CarcamoHernandez:2010im} 
  A.~E.~Carcamo Hernandez and R.~Rahman,
  Rev.\ Mex.\ Fis.\  {\bf 62}, no. 2, 100 (2016)
  [arXiv:1007.0447 [hep-ph]].



\bibitem{Branco:2010tx} 
  G.~C.~Branco, D.~Emmanuel-Costa and C.~Simoes,
  Phys.\ Lett.\ B {\bf 690}, 62 (2010)
  doi:10.1016/j.physletb.2010.05.009
  [arXiv:1001.5065 [hep-ph]].



\bibitem{Leser:2011fz} 
  P.~Leser and H.~Pas,
  Phys.\ Rev.\ D {\bf 84}, 017303 (2011)
  doi:10.1103/PhysRevD.84.017303
  [arXiv:1104.2448 [hep-ph]].



\bibitem{Gupta:2012dma} 
  M.~Gupta and G.~Ahuja,
  Int.\ J.\ Mod.\ Phys.\ A {\bf 27}, 1230033 (2012)
  doi:10.1142/S0217751X12300335
  [arXiv:1302.4823 [hep-ph]].



\bibitem{CarcamoHernandez:2012xy} 
  A.~E.~Carcamo Hernandez, C.~O.~Dib, N.~Neill H and A.~R.~Zerwekh,
  JHEP {\bf 1202}, 132 (2012)
  doi:10.1007/JHEP02(2012)132
  [arXiv:1201.0878 [hep-ph]].



\bibitem{Hernandez:2013mcf} 
  A.~E.~Carcamo Hernandez, R.~Martinez and F.~Ochoa,
  Phys.\ Rev.\ D {\bf 87}, no. 7, 075009 (2013)
  doi:10.1103/PhysRevD.87.075009
  [arXiv:1302.1757 [hep-ph]].



\bibitem{Pas:2014bra} 
  H.~Päs and E.~Schumacher,
  Phys.\ Rev.\ D {\bf 89}, no. 9, 096010 (2014)
  doi:10.1103/PhysRevD.89.096010
  [arXiv:1401.2328 [hep-ph]].



\bibitem{Hernandez:2014hka} 
  A.~E.~Carcamo Hernandez, S.~Kovalenko and I.~Schmidt,
  arXiv:1411.2913 [hep-ph].



\bibitem{Hernandez:2014zsa} 
  A.~E.~Cárcamo Hernández and I.~de Medeiros Varzielas,
  J.\ Phys.\ G {\bf 42}, no. 6, 065002 (2015)
  doi:10.1088/0954-3899/42/6/065002
  [arXiv:1410.2481 [hep-ph]].



\bibitem{Nishiura:2014psa} 
  H.~Nishiura and T.~Fukuyama,
  Mod.\ Phys.\ Lett.\ A {\bf 29}, 0147 (2014)
  doi:10.1142/S0217732314501478
  [arXiv:1405.2416 [hep-ph]].



\bibitem{Frank:2014aca} 
  M.~Frank, C.~Hamzaoui, N.~Pourtolami and M.~Toharia,
  Phys.\ Lett.\ B {\bf 742}, 178 (2015)
  doi:10.1016/j.physletb.2015.01.025
  [arXiv:1406.2331 [hep-ph]].



\bibitem{Ghosal:2015lwa} 
  A.~Ghosal and R.~Samanta,
  JHEP {\bf 1505}, 077 (2015)
  doi:10.1007/JHEP05(2015)077
  [arXiv:1501.00916 [hep-ph]].



\bibitem{Sinha:2015ooa} 
  R.~Sinha, R.~Samanta and A.~Ghosal,
  Phys.\ Lett.\ B {\bf 759}, 206 (2016)
  doi:10.1016/j.physletb.2016.05.080
  [arXiv:1508.05227 [hep-ph]].



\bibitem{Nishiura:2015qia} 
  H.~Nishiura and T.~Fukuyama,
  Phys.\ Lett.\ B {\bf 753}, 57 (2016)
  doi:10.1016/j.physletb.2015.11.080
  [arXiv:1510.01035 [hep-ph]].



\bibitem{Samanta:2015oqa} 
  R.~Samanta and A.~Ghosal,
  arXiv:1507.02582 [hep-ph].



\bibitem{Gautam:2015kya} 
  R.~R.~Gautam, M.~Singh and M.~Gupta,
  Phys.\ Rev.\ D {\bf 92}, no. 1, 013006 (2015)
  doi:10.1103/PhysRevD.92.013006
  [arXiv:1506.04868 [hep-ph]].



\bibitem{Pas:2015hca} 
  H.~Päs and E.~Schumacher,
  Phys.\ Rev.\ D {\bf 92}, no. 11, 114025 (2015)
  doi:10.1103/PhysRevD.92.114025
  [arXiv:1510.08757 [hep-ph]].



\bibitem{Hernandez:2015hrt} 
  A.~E.~Cárcamo Hernández,
  arXiv:1512.09092 [hep-ph].



\bibitem{Ishimori:2010au} 
  H.~Ishimori, T.~Kobayashi, H.~Ohki, Y.~Shimizu, H.~Okada and M.~Tanimoto,
  Prog.\ Theor.\ Phys.\ Suppl.\  {\bf 183}, 1 (2010)
  doi:10.1143/PTPS.183.1
  [arXiv:1003.3552 [hep-th]].



\bibitem{Altarelli:2010gt} 
  G.~Altarelli and F.~Feruglio,
  Rev.\ Mod.\ Phys.\  {\bf 82}, 2701 (2010)
  doi:10.1103/RevModPhys.82.2701
  [arXiv:1002.0211 [hep-ph]].



\bibitem{King:2013eh} 
  S.~F.~King and C.~Luhn,
  Rept.\ Prog.\ Phys.\  {\bf 76}, 056201 (2013)
  doi:10.1088/0034-4885/76/5/056201
  [arXiv:1301.1340 [hep-ph]].



\bibitem{King:2014nza} 
  S.~F.~King, A.~Merle, S.~Morisi, Y.~Shimizu and M.~Tanimoto,
  New J.\ Phys.\  {\bf 16}, 045018 (2014)
  doi:10.1088/1367-2630/16/4/045018
  [arXiv:1402.4271 [hep-ph]].



\bibitem{Ma:2001dn} 
  E.~Ma and G.~Rajasekaran,
  Phys.\ Rev.\ D {\bf 64}, 113012 (2001)
  doi:10.1103/PhysRevD.64.113012
  [hep-ph/0106291].



\bibitem{He:2006dk} 
  X.~G.~He, Y.~Y.~Keum and R.~R.~Volkas,
  JHEP {\bf 0604}, 039 (2006)
  doi:10.1088/1126-6708/2006/04/039
  [hep-ph/0601001].



\bibitem{Chen:2009um} 
  M.~C.~Chen and S.~F.~King,
  JHEP {\bf 0906}, 072 (2009)
  doi:10.1088/1126-6708/2009/06/072
  [arXiv:0903.0125 [hep-ph]].



\bibitem{Ahn:2012tv} 
  Y.~H.~Ahn and S.~K.~Kang,
  Phys.\ Rev.\ D {\bf 86}, 093003 (2012)
  doi:10.1103/PhysRevD.86.093003
  [arXiv:1203.4185 [hep-ph]].



\bibitem{Memenga:2013vc} 
  N.~Memenga, W.~Rodejohann and H.~Zhang,
  Phys.\ Rev.\ D {\bf 87}, no. 5, 053021 (2013)
  doi:10.1103/PhysRevD.87.053021
  [arXiv:1301.2963 [hep-ph]].



\bibitem{Felipe:2013vwa} 
  R.~Gonzalez Felipe, H.~Serodio and J.~P.~Silva,
  Phys.\ Rev.\ D {\bf 88}, no. 1, 015015 (2013)
  doi:10.1103/PhysRevD.88.015015
  [arXiv:1304.3468 [hep-ph]].



\bibitem{Varzielas:2012ai} 
  I.~de Medeiros Varzielas and D.~Pidt,
  JHEP {\bf 1303}, 065 (2013)
  doi:10.1007/JHEP03(2013)065
  [arXiv:1211.5370 [hep-ph]].



\bibitem{Ishimori:2012fg} 
  H.~Ishimori and E.~Ma,
  Phys.\ Rev.\ D {\bf 86}, 045030 (2012)
  doi:10.1103/PhysRevD.86.045030
  [arXiv:1205.0075 [hep-ph]].



\bibitem{King:2013hj} 
  S.~F.~King, S.~Morisi, E.~Peinado and J.~W.~F.~Valle,
  Phys.\ Lett.\ B {\bf 724}, 68 (2013)
  doi:10.1016/j.physletb.2013.05.067
  [arXiv:1301.7065 [hep-ph]].



\bibitem{Hernandez:2013dta} 
  A.~E.~Carcamo Hernandez, I.~de Medeiros Varzielas, S.~G.~Kovalenko, H.~Päs and I.~Schmidt,
  Phys.\ Rev.\ D {\bf 88}, no. 7, 076014 (2013)
  doi:10.1103/PhysRevD.88.076014
  [arXiv:1307.6499 [hep-ph]].



\bibitem{Babu:2002dz} 
  K.~S.~Babu, E.~Ma and J.~W.~F.~Valle,
  Phys.\ Lett.\ B {\bf 552}, 207 (2003)
  doi:10.1016/S0370-2693(02)03153-2
  [hep-ph/0206292].



\bibitem{Altarelli:2005yx} 
  G.~Altarelli and F.~Feruglio,
  Nucl.\ Phys.\ B {\bf 741}, 215 (2006)
  doi:10.1016/j.nuclphysb.2006.02.015
  [hep-ph/0512103].



\bibitem{Morisi:2013eca} 
  S.~Morisi, M.~Nebot, K.~M.~Patel, E.~Peinado and J.~W.~F.~Valle,
  Phys.\ Rev.\ D {\bf 88}, 036001 (2013)
  doi:10.1103/PhysRevD.88.036001
  [arXiv:1303.4394 [hep-ph]].



\bibitem{Altarelli:2005yp} 
  G.~Altarelli and F.~Feruglio,
  Nucl.\ Phys.\ B {\bf 720}, 64 (2005)
  doi:10.1016/j.nuclphysb.2005.05.005
  [hep-ph/0504165].



\bibitem{Kadosh:2010rm} 
  A.~Kadosh and E.~Pallante,
  JHEP {\bf 1008}, 115 (2010)
  doi:10.1007/JHEP08(2010)115
  [arXiv:1004.0321 [hep-ph]].



\bibitem{Kadosh:2013nra} 
  A.~Kadosh,
  JHEP {\bf 1306}, 114 (2013)
  doi:10.1007/JHEP06(2013)114
  [arXiv:1303.2645 [hep-ph]].



\bibitem{delAguila:2010vg} 
  F.~del Aguila, A.~Carmona and J.~Santiago,
  JHEP {\bf 1008}, 127 (2010)
  doi:10.1007/JHEP08(2010)127
  [arXiv:1001.5151 [hep-ph]].



\bibitem{Campos:2014lla} 
  M.~D.~Campos, A.~E.~Cárcamo Hernández, S.~Kovalenko, I.~Schmidt and E.~Schumacher,
  Phys.\ Rev.\ D {\bf 90}, no. 1, 016006 (2014)
  doi:10.1103/PhysRevD.90.016006
  [arXiv:1403.2525 [hep-ph]].



\bibitem{Vien:2014pta} 
  V.~V.~Vien and H.~N.~Long,
  Int.\ J.\ Mod.\ Phys.\ A {\bf 30}, no. 21, 1550117 (2015)
  doi:10.1142/S0217751X15501171
  [arXiv:1405.4665 [hep-ph]].



\bibitem{Hernandez:2015tna} 
  A.~E.~Cárcamo Hernández and R.~Martinez,
  Nucl.\ Phys.\ B {\bf 905}, 337 (2016)
  doi:10.1016/j.nuclphysb.2016.02.025
  [arXiv:1501.05937 [hep-ph]].



\bibitem{Kubo:2003pd} 
  J.~Kubo,
  Phys.\ Lett.\ B {\bf 578}, 156 (2004)
  Erratum: [Phys.\ Lett.\ B {\bf 619}, 387 (2005)]
  doi:10.1016/j.physletb.2005.06.013, 10.1016/j.physletb.2003.10.048
  [hep-ph/0309167].



\bibitem{Kobayashi:2003fh} 
  T.~Kobayashi, J.~Kubo and H.~Terao,
  Phys.\ Lett.\ B {\bf 568}, 83 (2003)
  doi:10.1016/j.physletb.2003.03.002
  [hep-ph/0303084].



\bibitem{Chen:2004rr} 
  S.~L.~Chen, M.~Frigerio and E.~Ma,
  Phys.\ Rev.\ D {\bf 70}, 073008 (2004)
  Erratum: [Phys.\ Rev.\ D {\bf 70}, 079905 (2004)]
  doi:10.1103/PhysRevD.70.079905, 10.1103/PhysRevD.70.073008
  [hep-ph/0404084].



\bibitem{Mondragon:2007af} 
  A.~Mondragon, M.~Mondragon and E.~Peinado,
  Phys.\ Rev.\ D {\bf 76}, 076003 (2007)
  doi:10.1103/PhysRevD.76.076003
  [arXiv:0706.0354 [hep-ph]].



\bibitem{Mondragon:2008gm} 
  A.~Mondragon, M.~Mondragon and E.~Peinado,
  Rev.\ Mex.\ Fis.\  {\bf 54}, no. 3, 81 (2008)
  [Rev.\ Mex.\ Fis.\ Suppl.\  {\bf 54}, 0181 (2008)]
  [arXiv:0805.3507 [hep-ph]].



\bibitem{Bhattacharyya:2010hp} 
  G.~Bhattacharyya, P.~Leser and H.~Pas,
  Phys.\ Rev.\ D {\bf 83}, 011701 (2011)
  doi:10.1103/PhysRevD.83.011701
  [arXiv:1006.5597 [hep-ph]].



\bibitem{Dong:2011vb} 
  P.~V.~Dong, H.~N.~Long, C.~H.~Nam and V.~V.~Vien,
  Phys.\ Rev.\ D {\bf 85}, 053001 (2012)
  doi:10.1103/PhysRevD.85.053001
  [arXiv:1111.6360 [hep-ph]].



\bibitem{Dias:2012bh} 
  A.~G.~Dias, A.~C.~B.~Machado and C.~C.~Nishi,
  Phys.\ Rev.\ D {\bf 86}, 093005 (2012)
  doi:10.1103/PhysRevD.86.093005
  [arXiv:1206.6362 [hep-ph]].



\bibitem{Meloni:2012ci} 
  D.~Meloni,
  JHEP {\bf 1205}, 124 (2012)
  doi:10.1007/JHEP05(2012)124
  [arXiv:1203.3126 [hep-ph]].



\bibitem{Canales:2012dr} 
  F.~Gonzalez Canales, A.~Mondragon and M.~Mondragon,
  Fortsch.\ Phys.\  {\bf 61}, 546 (2013)
  doi:10.1002/prop.201200121
  [arXiv:1205.4755 [hep-ph]].



\bibitem{Canales:2013cga} 
  F.~González Canales, A.~Mondragón, M.~Mondragón, U.~J.~Saldaña Salazar and L.~Velasco-Sevilla,
  Phys.\ Rev.\ D {\bf 88}, 096004 (2013)
  doi:10.1103/PhysRevD.88.096004
  [arXiv:1304.6644 [hep-ph]].



\bibitem{Ma:2013zca} 
  E.~Ma and B.~Melic,
  Phys.\ Lett.\ B {\bf 725}, 402 (2013)
  doi:10.1016/j.physletb.2013.07.015
  [arXiv:1303.6928 [hep-ph]].



\bibitem{Kajiyama:2013sza} 
  Y.~Kajiyama, H.~Okada and K.~Yagyu,
  Nucl.\ Phys.\ B {\bf 887}, 358 (2014)
  doi:10.1016/j.nuclphysb.2014.08.009
  [arXiv:1309.6234 [hep-ph]].



\bibitem{Hernandez:2013hea} 
  A.~E.~Cárcamo Hernández, R.~Martínez and F.~Ochoa,
  arXiv:1309.6567 [hep-ph].



\bibitem{Ma:2014qra} 
  E.~Ma and R.~Srivastava,
  Phys.\ Lett.\ B {\bf 741}, 217 (2015)
  doi:10.1016/j.physletb.2014.12.049
  [arXiv:1411.5042 [hep-ph]].



\bibitem{Hernandez:2014vta} 
  A.~E.~Cárcamo Hernández, R.~Martinez and J.~Nisperuza,
  Eur.\ Phys.\ J.\ C {\bf 75}, no. 2, 72 (2015)
  doi:10.1140/epjc/s10052-015-3278-z
  [arXiv:1401.0937 [hep-ph]].



\bibitem{Hernandez:2014lpa} 
  A.~E.~Cárcamo Hernández, E.~Cataño Mur and R.~Martinez,
  Phys.\ Rev.\ D {\bf 90}, no. 7, 073001 (2014)
  doi:10.1103/PhysRevD.90.073001
  [arXiv:1407.5217 [hep-ph]].



\bibitem{Gupta:2014nba} 
  S.~Gupta, C.~S.~Kim and P.~Sharma,
  Phys.\ Lett.\ B {\bf 740}, 353 (2015)
  doi:10.1016/j.physletb.2014.12.005
  [arXiv:1408.0172 [hep-ph]].



\bibitem{Hernandez:2015dga} 
  A.~E.~Cárcamo Hernández, I.~de Medeiros Varzielas and E.~Schumacher,
  Phys.\ Rev.\ D {\bf 93}, no. 1, 016003 (2016)
  doi:10.1103/PhysRevD.93.016003
  [arXiv:1509.02083 [hep-ph]].



\bibitem{Hernandez:2015zeh} 
  A.~E.~Cárcamo Hernández, I.~de Medeiros Varzielas and N.~A.~Neill,
  arXiv:1511.07420 [hep-ph].



\bibitem{Hernandez:2016rbi} 
  A.~E.~Cárcamo Hernández, I.~de Medeiros Varzielas and E.~Schumacher,
  arXiv:1601.00661 [hep-ph].



\bibitem{Mohapatra:2012tb} 
  R.~N.~Mohapatra and C.~C.~Nishi,
  Phys.\ Rev.\ D {\bf 86}, 073007 (2012)
  doi:10.1103/PhysRevD.86.073007
  [arXiv:1208.2875 [hep-ph]].



\bibitem{BhupalDev:2012nm} 
  P.~S.~Bhupal Dev, B.~Dutta, R.~N.~Mohapatra and M.~Severson,
  Phys.\ Rev.\ D {\bf 86}, 035002 (2012)
  doi:10.1103/PhysRevD.86.035002
  [arXiv:1202.4012 [hep-ph]].



\bibitem{Varzielas:2012pa} 
  I.~de Medeiros Varzielas and L.~Lavoura,
  J.\ Phys.\ G {\bf 40}, 085002 (2013)
  doi:10.1088/0954-3899/40/8/085002
  [arXiv:1212.3247 [hep-ph]].



\bibitem{Ding:2013hpa} 
  G.~J.~Ding, S.~F.~King, C.~Luhn and A.~J.~Stuart,
  JHEP {\bf 1305}, 084 (2013)
  doi:10.1007/JHEP05(2013)084
  [arXiv:1303.6180 [hep-ph]].



\bibitem{Ishimori:2010fs} 
  H.~Ishimori, Y.~Shimizu, M.~Tanimoto and A.~Watanabe,
  Phys.\ Rev.\ D {\bf 83}, 033004 (2011)
  doi:10.1103/PhysRevD.83.033004
  [arXiv:1010.3805 [hep-ph]].



\bibitem{Ding:2013eca} 
  G.~J.~Ding and Y.~L.~Zhou,
  Nucl.\ Phys.\ B {\bf 876}, 418 (2013)
  doi:10.1016/j.nuclphysb.2013.08.011
  [arXiv:1304.2645 [hep-ph]].



\bibitem{Hagedorn:2011un} 
  C.~Hagedorn and M.~Serone,
  JHEP {\bf 1110}, 083 (2011)
  doi:10.1007/JHEP10(2011)083
  [arXiv:1106.4021 [hep-ph]].



\bibitem{Campos:2014zaa} 
  M.~D.~Campos, A.~E.~Cárcamo Hernández, H.~Päs and E.~Schumacher,
  Phys.\ Rev.\ D {\bf 91}, no. 11, 116011 (2015)
  doi:10.1103/PhysRevD.91.116011
  [arXiv:1408.1652 [hep-ph]].



\bibitem{Dong:2010zu} 
  P.~V.~Dong, H.~N.~Long, D.~V.~Soa and V.~V.~Vien,
  Eur.\ Phys.\ J.\ C {\bf 71}, 1544 (2011)
  doi:10.1140/epjc/s10052-011-1544-2
  [arXiv:1009.2328 [hep-ph]].



\bibitem{VanVien:2015xha} 
  V.~V.~Vien, H.~N.~Long and D.~P.~Khoi,
  Int.\ J.\ Mod.\ Phys.\ A {\bf 30}, no. 17, 1550102 (2015)
  doi:10.1142/S0217751X1550102X
  [arXiv:1506.06063 [hep-ph]].



\bibitem{Arbelaez:2016mhg} 
  C.~Arbeláez, A.~E.~Cárcamo Hernández, S.~Kovalenko and I.~Schmidt,
  arXiv:1602.03607 [hep-ph].



\bibitem{Frampton:1994rk} 
  P.~H.~Frampton and T.~W.~Kephart,
  Int.\ J.\ Mod.\ Phys.\ A {\bf 10}, 4689 (1995)
  doi:10.1142/S0217751X95002187
  [hep-ph/9409330].



\bibitem{Grimus:2003kq} 
  W.~Grimus and L.~Lavoura,
  Phys.\ Lett.\ B {\bf 572}, 189 (2003)
  doi:10.1016/j.physletb.2003.08.032
  [hep-ph/0305046].



\bibitem{Grimus:2004rj} 
  W.~Grimus, A.~S.~Joshipura, S.~Kaneko, L.~Lavoura and M.~Tanimoto,
  JHEP {\bf 0407}, 078 (2004)
  doi:10.1088/1126-6708/2004/07/078
  [hep-ph/0407112].



\bibitem{Frigerio:2004jg} 
  M.~Frigerio, S.~Kaneko, E.~Ma and M.~Tanimoto,
  Phys.\ Rev.\ D {\bf 71}, 011901 (2005)
  doi:10.1103/PhysRevD.71.011901
  [hep-ph/0409187].



\bibitem{Babu:2004tn} 
  K.~S.~Babu and J.~Kubo,
  Phys.\ Rev.\ D {\bf 71}, 056006 (2005)
  doi:10.1103/PhysRevD.71.056006
  [hep-ph/0411226].



\bibitem{Adulpravitchai:2008yp} 
  A.~Adulpravitchai, A.~Blum and C.~Hagedorn,
  JHEP {\bf 0903}, 046 (2009)
  doi:10.1088/1126-6708/2009/03/046
  [arXiv:0812.3799 [hep-ph]].



\bibitem{Ishimori:2008gp} 
  H.~Ishimori, T.~Kobayashi, H.~Ohki, Y.~Omura, R.~Takahashi and M.~Tanimoto,
  Phys.\ Lett.\ B {\bf 662}, 178 (2008)
  doi:10.1016/j.physletb.2008.03.007
  [arXiv:0802.2310 [hep-ph]].



\bibitem{Hagedorn:2010mq} 
  C.~Hagedorn and R.~Ziegler,
  Phys.\ Rev.\ D {\bf 82}, 053011 (2010)
  doi:10.1103/PhysRevD.82.053011
  [arXiv:1007.1888 [hep-ph]].



\bibitem{Meloni:2011cc} 
  D.~Meloni, S.~Morisi and E.~Peinado,
  Phys.\ Lett.\ B {\bf 703}, 281 (2011)
  doi:10.1016/j.physletb.2011.07.084
  [arXiv:1104.0178 [hep-ph]].



\bibitem{Vien:2013zra} 
  V.~V.~Vien and H.~N.~Long,
  Int.\ J.\ Mod.\ Phys.\ A {\bf 28}, 1350159 (2013)
  doi:10.1142/S0217751X13501595
  [arXiv:1312.5034 [hep-ph]].



\bibitem{Kawashima:2009jv} 
  K.~Kawashima, J.~Kubo and A.~Lenz,
  Phys.\ Lett.\ B {\bf 681}, 60 (2009)
  doi:10.1016/j.physletb.2009.09.064
  [arXiv:0907.2302 [hep-ph]].



\bibitem{Kaburaki:2010xc} 
  Y.~Kaburaki, K.~Konya, J.~Kubo and A.~Lenz,
  Phys.\ Rev.\ D {\bf 84}, 016007 (2011)
  doi:10.1103/PhysRevD.84.016007
  [arXiv:1012.2435 [hep-ph]].



\bibitem{Babu:2011mv} 
  K.~S.~Babu, K.~Kawashima and J.~Kubo,
  Phys.\ Rev.\ D {\bf 83}, 095008 (2011)
  doi:10.1103/PhysRevD.83.095008
  [arXiv:1103.1664 [hep-ph]].



\bibitem{Gomez-Izquierdo:2013uaa} 
  J.~C.~Gómez-Izquierdo, F.~González-Canales and M.~Mondragon,
  Eur.\ Phys.\ J.\ C {\bf 75}, no. 5, 221 (2015)
  doi:10.1140/epjc/s10052-015-3440-7
  [arXiv:1312.7385 [hep-ph]].



\bibitem{Luhn:2007sy} 
  C.~Luhn, S.~Nasri and P.~Ramond,
  Phys.\ Lett.\ B {\bf 652}, 27 (2007)
  doi:10.1016/j.physletb.2007.06.059
  [arXiv:0706.2341 [hep-ph]].



\bibitem{Hagedorn:2008bc} 
  C.~Hagedorn, M.~A.~Schmidt and A.~Y.~Smirnov,
  Phys.\ Rev.\ D {\bf 79}, 036002 (2009)
  doi:10.1103/PhysRevD.79.036002
  [arXiv:0811.2955 [hep-ph]].



\bibitem{Cao:2010mp} 
  Q.~H.~Cao, S.~Khalil, E.~Ma and H.~Okada,
  Phys.\ Rev.\ Lett.\  {\bf 106}, 131801 (2011)
  doi:10.1103/PhysRevLett.106.131801
  [arXiv:1009.5415 [hep-ph]].



\bibitem{Luhn:2012bc} 
  C.~Luhn, K.~M.~Parattu and A.~Wingerter,
  JHEP {\bf 1212}, 096 (2012)
  doi:10.1007/JHEP12(2012)096
  [arXiv:1210.1197 [hep-ph]].



\bibitem{Kajiyama:2013lja} 
  Y.~Kajiyama, H.~Okada and K.~Yagyu,
  JHEP {\bf 1310}, 196 (2013)
  doi:10.1007/JHEP10(2013)196
  [arXiv:1307.0480 [hep-ph]].



\bibitem{Bonilla:2014xla} 
  C.~Bonilla, S.~Morisi, E.~Peinado and J.~W.~F.~Valle,
  Phys.\ Lett.\ B {\bf 742}, 99 (2015)
  doi:10.1016/j.physletb.2015.01.017
  [arXiv:1411.4883 [hep-ph]].



\bibitem{Vien:2014gza} 
  V.~V.~Vien and H.~N.~Long,
  JHEP {\bf 1404}, 133 (2014)
  doi:10.1007/JHEP04(2014)133
  [arXiv:1402.1256 [hep-ph]].



\bibitem{Vien:2015koa} 
  V.~V.~Vien,
  Mod.\ Phys.\ Lett.\ A {\bf 29}, 28 (2014)
  doi:10.1142/S0217732314501399
  [arXiv:1508.02585 [hep-ph]].



\bibitem{Hernandez:2015cra} 
  A.~E.~Cárcamo Hernández and R.~Martinez,
  J.\ Phys.\ G {\bf 43}, no. 4, 045003 (2016)
  doi:10.1088/0954-3899/43/4/045003
  [arXiv:1501.07261 [hep-ph]].



\bibitem{Arbelaez:2015toa} 
  C.~Arbeláez, A.~E.~Cárcamo Hernández, S.~Kovalenko and I.~Schmidt,
  Phys.\ Rev.\ D {\bf 92}, no. 11, 115015 (2015)
  doi:10.1103/PhysRevD.92.115015
  [arXiv:1507.03852 [hep-ph]].



\bibitem{Ding:2011qt} 
  G.~J.~Ding,
  Nucl.\ Phys.\ B {\bf 853}, 635 (2011)
  doi:10.1016/j.nuclphysb.2011.08.012
  [arXiv:1105.5879 [hep-ph]].



\bibitem{Hartmann:2011dn} 
  C.~Hartmann,
  Phys.\ Rev.\ D {\bf 85}, 013012 (2012)
  doi:10.1103/PhysRevD.85.013012
  [arXiv:1109.5143 [hep-ph]].



\bibitem{Hartmann:2011pq} 
  C.~Hartmann and A.~Zee,
  Nucl.\ Phys.\ B {\bf 853}, 105 (2011)
  doi:10.1016/j.nuclphysb.2011.07.023
  [arXiv:1106.0333 [hep-ph]].



\bibitem{Kajiyama:2010sb} 
  Y.~Kajiyama and H.~Okada,
  Nucl.\ Phys.\ B {\bf 848}, 303 (2011)
  doi:10.1016/j.nuclphysb.2011.02.020
  [arXiv:1011.5753 [hep-ph]].



\bibitem{Aranda:2000tm} 
  A.~Aranda, C.~D.~Carone and R.~F.~Lebed,
  Phys.\ Rev.\ D {\bf 62}, 016009 (2000)
  doi:10.1103/PhysRevD.62.016009
  [hep-ph/0002044].



\bibitem{Aranda:2007dp} 
  A.~Aranda,
  Phys.\ Rev.\ D {\bf 76}, 111301 (2007)
  doi:10.1103/PhysRevD.76.111301
  [arXiv:0707.3661 [hep-ph]].



\bibitem{Chen:2007afa} 
  M.~C.~Chen and K.~T.~Mahanthappa,
  Phys.\ Lett.\ B {\bf 652}, 34 (2007)
  doi:10.1016/j.physletb.2007.06.064
  [arXiv:0705.0714 [hep-ph]].



\bibitem{Frampton:2008bz} 
  P.~H.~Frampton, T.~W.~Kephart and S.~Matsuzaki,
  Phys.\ Rev.\ D {\bf 78}, 073004 (2008)
  doi:10.1103/PhysRevD.78.073004
  [arXiv:0807.4713 [hep-ph]].



\bibitem{Eby:2011ph} 
  D.~A.~Eby, P.~H.~Frampton, X.~G.~He and T.~W.~Kephart,
  Phys.\ Rev.\ D {\bf 84}, 037302 (2011)
  doi:10.1103/PhysRevD.84.037302
  [arXiv:1103.5737 [hep-ph]].



\bibitem{Frampton:2013lva} 
  P.~H.~Frampton, C.~M.~Ho and T.~W.~Kephart,
  Phys.\ Rev.\ D {\bf 89}, no. 2, 027701 (2014)
  doi:10.1103/PhysRevD.89.027701
  [arXiv:1305.4402 [hep-ph]].



\bibitem{Chen:2013wba} 
  M.~C.~Chen, J.~Huang, K.~T.~Mahanthappa and A.~M.~Wijangco,
  JHEP {\bf 1310}, 112 (2013)
  doi:10.1007/JHEP10(2013)112
  [arXiv:1307.7711 [hep-ph]].



\bibitem{Ma:2007wu} 
  E.~Ma,
  Phys.\ Lett.\ B {\bf 660}, 505 (2008)
  doi:10.1016/j.physletb.2007.12.060
  [arXiv:0709.0507 [hep-ph]].



\bibitem{Varzielas:2012nn} 
  I.~de Medeiros Varzielas, D.~Emmanuel-Costa and P.~Leser,
  Phys.\ Lett.\ B {\bf 716}, 193 (2012)
  doi:10.1016/j.physletb.2012.08.008
  [arXiv:1204.3633 [hep-ph]].



\bibitem{Bhattacharyya:2012pi} 
  G.~Bhattacharyya, I.~de Medeiros Varzielas and P.~Leser,
  Phys.\ Rev.\ Lett.\  {\bf 109}, 241603 (2012)
  doi:10.1103/PhysRevLett.109.241603
  [arXiv:1210.0545 [hep-ph]].



\bibitem{Ma:2013xqa} 
  E.~Ma,
  Phys.\ Lett.\ B {\bf 723}, 161 (2013)
  doi:10.1016/j.physletb.2013.05.011
  [arXiv:1304.1603 [hep-ph]].



\bibitem{Nishi:2013jqa} 
  C.~C.~Nishi,
  Phys.\ Rev.\ D {\bf 88}, no. 3, 033010 (2013)
  doi:10.1103/PhysRevD.88.033010
  [arXiv:1306.0877 [hep-ph]].



\bibitem{Varzielas:2013sla} 
  I.~de Medeiros Varzielas and D.~Pidt,
  J.\ Phys.\ G {\bf 41}, 025004 (2014)
  doi:10.1088/0954-3899/41/2/025004
  [arXiv:1307.0711 [hep-ph]].



\bibitem{Aranda:2013gga} 
  A.~Aranda, C.~Bonilla, S.~Morisi, E.~Peinado and J.~W.~F.~Valle,
  Phys.\ Rev.\ D {\bf 89}, no. 3, 033001 (2014)
  doi:10.1103/PhysRevD.89.033001
  [arXiv:1307.3553 [hep-ph]].



\bibitem{Ma:2014eka} 
  E.~Ma and A.~Natale,
  Phys.\ Lett.\ B {\bf 734}, 403 (2014)
  doi:10.1016/j.physletb.2014.05.070
  [arXiv:1403.6772 [hep-ph]].



\bibitem{Abbas:2014ewa} 
  M.~Abbas and S.~Khalil,
  Phys.\ Rev.\ D {\bf 91}, no. 5, 053003 (2015)
  doi:10.1103/PhysRevD.91.053003
  [arXiv:1406.6716 [hep-ph]].



\bibitem{Abbas:2015zna} 
  M.~Abbas, S.~Khalil, A.~Rashed and A.~Sil,
  Phys.\ Rev.\ D {\bf 93}, no. 1, 013018 (2016)
  doi:10.1103/PhysRevD.93.013018
  [arXiv:1508.03727 [hep-ph]].



\bibitem{Varzielas:2015aua} 
  I.~de Medeiros Varzielas,
  JHEP {\bf 1508}, 157 (2015)
  doi:10.1007/JHEP08(2015)157
  [arXiv:1507.00338 [hep-ph]].



\bibitem{Bjorkeroth:2015uou} 
  F.~Björkeroth, F.~J.~de Anda, I.~de Medeiros Varzielas and S.~F.~King,
  Phys.\ Rev.\ D {\bf 94}, no. 1, 016006 (2016)
  doi:10.1103/PhysRevD.94.016006
  [arXiv:1512.00850 [hep-ph]].



\bibitem{Chen:2015jta} 
  P.~Chen, G.~J.~Ding, A.~D.~Rojas, C.~A.~Vaquera-Araujo and J.~W.~F.~Valle,
  JHEP {\bf 1601}, 007 (2016)
  doi:10.1007/JHEP01(2016)007
  [arXiv:1509.06683 [hep-ph]].



\bibitem{Vien:2016tmh} 
  V.~V.~Vien, A.~E.~Cárcamo Hernández and H.~N.~Long,
  arXiv:1601.03300 [hep-ph].



\bibitem{Hernandez:2016eod} 
  A.~E.~Cárcamo Hernández, H.~N.~Long and V.~V.~Vien,
  Eur.\ Phys.\ J.\ C {\bf 76}, no. 5, 242 (2016)
  doi:10.1140/epjc/s10052-016-4074-0
  [arXiv:1601.05062 [hep-ph]].



\bibitem{Everett:2008et} 
  L.~L.~Everett and A.~J.~Stuart,
  Phys.\ Rev.\ D {\bf 79}, 085005 (2009)
  doi:10.1103/PhysRevD.79.085005
  [arXiv:0812.1057 [hep-ph]].



\bibitem{Feruglio:2011qq} 
  F.~Feruglio and A.~Paris,
  JHEP {\bf 1103}, 101 (2011)
  doi:10.1007/JHEP03(2011)101
  [arXiv:1101.0393 [hep-ph]].



\bibitem{Cooper:2012bd} 
  I.~K.~Cooper, S.~F.~King and A.~J.~Stuart,
  Nucl.\ Phys.\ B {\bf 875}, 650 (2013)
  doi:10.1016/j.nuclphysb.2013.07.027
  [arXiv:1212.1066 [hep-ph]].



\bibitem{Varzielas:2013hga} 
  I.~de Medeiros Varzielas and L.~Lavoura,
  J.\ Phys.\ G {\bf 41}, 055005 (2014)
  doi:10.1088/0954-3899/41/5/055005
  [arXiv:1312.0215 [hep-ph]].



\bibitem{Gehrlein:2014wda} 
  J.~Gehrlein, J.~P.~Oppermann, D.~Schäfer and M.~Spinrath,
  Nucl.\ Phys.\ B {\bf 890}, 539 (2014)
  doi:10.1016/j.nuclphysb.2014.11.023
  [arXiv:1410.2057 [hep-ph]].



\bibitem{Gehrlein:2015dxa} 
  J.~Gehrlein, S.~T.~Petcov, M.~Spinrath and X.~Zhang,
  Nucl.\ Phys.\ B {\bf 896}, 311 (2015)
  doi:10.1016/j.nuclphysb.2015.04.019
  [arXiv:1502.00110 [hep-ph]].



\bibitem{DiIura:2015kfa} 
  A.~Di Iura, C.~Hagedorn and D.~Meloni,
  JHEP {\bf 1508}, 037 (2015)
  doi:10.1007/JHEP08(2015)037
  [arXiv:1503.04140 [hep-ph]].



\bibitem{Ballett:2015wia} 
  P.~Ballett, S.~Pascoli and J.~Turner,
  Phys.\ Rev.\ D {\bf 92}, no. 9, 093008 (2015)
  doi:10.1103/PhysRevD.92.093008
  [arXiv:1503.07543 [hep-ph]].



\bibitem{Gehrlein:2015dza} 
  J.~Gehrlein, S.~T.~Petcov, M.~Spinrath and X.~Zhang,
  Nucl.\ Phys.\ B {\bf 899}, 617 (2015)
  doi:10.1016/j.nuclphysb.2015.08.019
  [arXiv:1508.07930 [hep-ph]].



\bibitem{Turner:2015uta} 
  J.~Turner,
  Phys.\ Rev.\ D {\bf 92}, no. 11, 116007 (2015)
  doi:10.1103/PhysRevD.92.116007
  [arXiv:1507.06224 [hep-ph]].



\bibitem{Li:2015jxa} 
  C.~C.~Li and G.~J.~Ding,
  JHEP {\bf 1505}, 100 (2015)
  doi:10.1007/JHEP05(2015)100
  [arXiv:1503.03711 [hep-ph]].



\bibitem{Pakvasa:1977in} 
  S.~Pakvasa and H.~Sugawara,
  Phys.\ Lett.\ B {\bf 73}, 61 (1978).
  doi:10.1016/0370-2693(78)90172-7



\bibitem{Georgi:1978bv} 
  H.~Georgi and A.~Pais,
  Phys.\ Rev.\ D {\bf 19}, 2746 (1979).
  doi:10.1103/PhysRevD.19.2746



\bibitem{Valle:1983dk} 
  J.~W.~F.~Valle and M.~Singer,
  Phys.\ Rev.\ D {\bf 28}, 540 (1983).
  doi:10.1103/PhysRevD.28.540



\bibitem{Pisano:1991ee} 
  F.~Pisano and V.~Pleitez,
  Phys.\ Rev.\ D {\bf 46}, 410 (1992)
  doi:10.1103/PhysRevD.46.410
  [hep-ph/9206242].



\bibitem{Montero:1992jk} 
  J.~C.~Montero, F.~Pisano and V.~Pleitez,
  Phys.\ Rev.\ D {\bf 47}, 2918 (1993)
  doi:10.1103/PhysRevD.47.2918
  [hep-ph/9212271].



\bibitem{Foot:1992rh} 
  R.~Foot, O.~F.~Hernandez, F.~Pisano and V.~Pleitez,
  Phys.\ Rev.\ D {\bf 47}, 4158 (1993)
  doi:10.1103/PhysRevD.47.4158
  [hep-ph/9207264].



\bibitem{Frampton:1992wt} 
  P.~H.~Frampton,
  Phys.\ Rev.\ Lett.\  {\bf 69}, 2889 (1992).
  doi:10.1103/PhysRevLett.69.2889



\bibitem{Ng:1992st} 
  D.~Ng,
  Phys.\ Rev.\ D {\bf 49}, 4805 (1994)
  doi:10.1103/PhysRevD.49.4805
  [hep-ph/9212284].



\bibitem{Duong:1993zn} 
  T.~V.~Duong and E.~Ma,
  Phys.\ Lett.\ B {\bf 316}, 307 (1993)
  doi:10.1016/0370-2693(93)90329-G
  [hep-ph/9306264].



\bibitem{Hoang:1996gi} 
  H.~N.~Long,
  Phys.\ Rev.\ D {\bf 54}, 4691 (1996)
  doi:10.1103/PhysRevD.54.4691
  [hep-ph/9607439].



\bibitem{Hoang:1995vq} 
  H.~N.~Long,
  Phys.\ Rev.\ D {\bf 53}, 437 (1996)
  doi:10.1103/PhysRevD.53.437
  [hep-ph/9504274].



\bibitem{Foot:1994ym} 
  R.~Foot, H.~N.~Long and T.~A.~Tran,
  Phys.\ Rev.\ D {\bf 50}, no. 1, R34 (1994)
  doi:10.1103/PhysRevD.50.R34
  [hep-ph/9402243].



\bibitem{Martinez:2001mu} 
  R.~Martinez, W.~A.~Ponce and L.~A.~Sanchez,
  Phys.\ Rev.\ D {\bf 65}, 055013 (2002)
  doi:10.1103/PhysRevD.65.055013
  [hep-ph/0110246].



\bibitem{Sanchez:2001ua} 
  L.~A.~Sanchez, W.~A.~Ponce and R.~Martinez,
  Phys.\ Rev.\ D {\bf 64}, 075013 (2001)
  doi:10.1103/PhysRevD.64.075013
  [hep-ph/0103244].



\bibitem{Diaz:2003dk} 
  R.~A.~Diaz, R.~Martinez and F.~Ochoa,
  Phys.\ Rev.\ D {\bf 69}, 095009 (2004)
  doi:10.1103/PhysRevD.69.095009
  [hep-ph/0309280].



\bibitem{Diaz:2004fs} 
  R.~A.~Diaz, R.~Martinez and F.~Ochoa,
  Phys.\ Rev.\ D {\bf 72}, 035018 (2005)
  doi:10.1103/PhysRevD.72.035018
  [hep-ph/0411263].



\bibitem{Dias:2004dc} 
  A.~G.~Dias, R.~Martinez and V.~Pleitez,
  Eur.\ Phys.\ J.\ C {\bf 39}, 101 (2005)
  doi:10.1140/epjc/s2004-02083-0
  [hep-ph/0407141].



\bibitem{Dias:2005yh} 
  A.~G.~Dias, C.~A.~de S.Pires and P.~S.~Rodrigues da Silva,
  Phys.\ Lett.\ B {\bf 628}, 85 (2005)
  doi:10.1016/j.physletb.2005.09.028
  [hep-ph/0508186].



\bibitem{Dias:2005jm} 
  A.~G.~Dias, A.~Doff, C.~A.~de S.Pires and P.~S.~Rodrigues da Silva,
  Phys.\ Rev.\ D {\bf 72}, 035006 (2005)
  doi:10.1103/PhysRevD.72.035006
  [hep-ph/0503014].



\bibitem{Ochoa:2005ch} 
  F.~Ochoa and R.~Martinez,
  Phys.\ Rev.\ D {\bf 72}, 035010 (2005)
  doi:10.1103/PhysRevD.72.035010
  [hep-ph/0505027].



\bibitem{CarcamoHernandez:2005ka} 
  A.~E.~Carcamo Hernandez, R.~Martinez and F.~Ochoa,
  Phys.\ Rev.\ D {\bf 73}, 035007 (2006)
  doi:10.1103/PhysRevD.73.035007
  [hep-ph/0510421].



\bibitem{Salazar:2007ym} 
  J.~C.~Salazar, W.~A.~Ponce and D.~A.~Gutierrez,
  Phys.\ Rev.\ D {\bf 75}, 075016 (2007)
  doi:10.1103/PhysRevD.75.075016
  [hep-ph/0703300 [HEP-PH]].



\bibitem{Benavides:2009cn} 
  R.~H.~Benavides, Y.~Giraldo and W.~A.~Ponce,
  Phys.\ Rev.\ D {\bf 80}, 113009 (2009)
  doi:10.1103/PhysRevD.80.113009
  [arXiv:0911.3568 [hep-ph]].



\bibitem{Dias:2010vt} 
  A.~G.~Dias, C.~A.~de S.Pires and P.~S.~Rodrigues da Silva,
  Phys.\ Rev.\ D {\bf 82}, 035013 (2010)
  doi:10.1103/PhysRevD.82.035013
  [arXiv:1003.3260 [hep-ph]].



\bibitem{Dias:2012xp} 
  A.~G.~Dias, C.~A.~de S.Pires, P.~S.~Rodrigues da Silva and A.~Sampieri,
  Phys.\ Rev.\ D {\bf 86}, 035007 (2012)
  doi:10.1103/PhysRevD.86.035007
  [arXiv:1206.2590].



\bibitem{Alvarado:2012xi} 
  C.~Alvarado, R.~Martinez and F.~Ochoa,
  Phys.\ Rev.\ D {\bf 86}, 025027 (2012)
  doi:10.1103/PhysRevD.86.025027
  [arXiv:1207.0014 [hep-ph]].



\bibitem{Catano:2012kw} 
  M.~E.~Catano, R.~Martinez and F.~Ochoa,
  Phys.\ Rev.\ D {\bf 86}, 073015 (2012)
  doi:10.1103/PhysRevD.86.073015
  [arXiv:1206.1966 [hep-ph]].



\bibitem{Boucenna:2014ela} 
  S.~M.~Boucenna, S.~Morisi and J.~W.~F.~Valle,
  Phys.\ Rev.\ D {\bf 90}, no. 1, 013005 (2014)
  doi:10.1103/PhysRevD.90.013005
  [arXiv:1405.2332 [hep-ph]].



\bibitem{Boucenna:2014dia} 
  S.~M.~Boucenna, R.~M.~Fonseca, F.~Gonzalez-Canales and J.~W.~F.~Valle,
  Phys.\ Rev.\ D {\bf 91}, no. 3, 031702 (2015)
  doi:10.1103/PhysRevD.91.031702
  [arXiv:1411.0566 [hep-ph]].



\bibitem{Phong:2014ofa} 
  V.~Q.~Phong, H.~N.~Long, V.~T.~Van and L.~H.~Minh,
  Eur.\ Phys.\ J.\ C {\bf 75}, no. 7, 342 (2015)
  doi:10.1140/epjc/s10052-015-3550-2
  [arXiv:1409.0750 [hep-ph]].



\bibitem{Boucenna:2015zwa} 
  S.~M.~Boucenna, J.~W.~F.~Valle and A.~Vicente,
  Phys.\ Rev.\ D {\bf 92}, no. 5, 053001 (2015)
  doi:10.1103/PhysRevD.92.053001
  [arXiv:1502.07546 [hep-ph]].



\bibitem{DeConto:2015eia} 
  G.~De Conto, A.~C.~B.~Machado and V.~Pleitez,
  Phys.\ Rev.\ D {\bf 92}, no. 7, 075031 (2015)
  doi:10.1103/PhysRevD.92.075031
  [arXiv:1505.01343 [hep-ph]].



\bibitem{Correia:2015tra} 
  F.~C.~Correia and V.~Pleitez,
  Phys.\ Rev.\ D {\bf 92}, 113006 (2015)
  doi:10.1103/PhysRevD.92.113006
  [arXiv:1508.07319 [hep-ph]].



\bibitem{Dong:2015rka} 
  P.~V.~Dong, C.~S.~Kim, D.~V.~Soa and N.~T.~Thuy,
  Phys.\ Rev.\ D {\bf 91}, no. 11, 115019 (2015)
  doi:10.1103/PhysRevD.91.115019
  [arXiv:1501.04385 [hep-ph]].



\bibitem{Okada:2015bxa} 
  H.~Okada, N.~Okada and Y.~Orikasa,
  Phys.\ Rev.\ D {\bf 93}, no. 7, 073006 (2016)
  doi:10.1103/PhysRevD.93.073006
  [arXiv:1504.01204 [hep-ph]].



\bibitem{Binh:2015cba} 
  D.~T.~Binh, D.~T.~Huong and H.~N.~Long,
  Zh.\ Eksp.\ Teor.\ Fiz.\  {\bf 148}, 1115 (2015)
  [J.\ Exp.\ Theor.\ Phys.\  {\bf 121}, no. 6, 976 (2015)]
  doi:10.7868/S004445101512007X, 10.1134/S1063776115120109
  [arXiv:1504.03510 [hep-ph]].



\bibitem{Hue:2015fbb} 
  L.~T.~Hue, H.~N.~Long, T.~T.~Thuc and T.~Phong Nguyen,
  Nucl.\ Phys.\ B {\bf 907}, 37 (2016)
  doi:10.1016/j.nuclphysb.2016.03.034
  [arXiv:1512.03266 [hep-ph]].



\bibitem{Benavides:2015afa} 
  R.~H.~Benavides, L.~N.~Epele, H.~Fanchiotti, C.~G.~Canal and W.~A.~Ponce,
  Adv.\ High Energy Phys.\  {\bf 2015}, 813129 (2015)
  doi:10.1155/2015/813129
  [arXiv:1503.01686 [hep-ph]].



\bibitem{Boucenna:2015pav} 
  S.~M.~Boucenna, S.~Morisi and A.~Vicente,
  Phys.\ Rev.\ D {\bf 93}, no. 11, 115008 (2016)
  doi:10.1103/PhysRevD.93.115008
  [arXiv:1512.06878 [hep-ph]].



\bibitem{Hernandez:2015ywg} 
  A.~E.~C.~Hernández and I.~Nišandžić,
  Eur.\ Phys.\ J.\ C {\bf 76}, no. 7, 380 (2016)
  doi:10.1140/epjc/s10052-016-4230-6
  [arXiv:1512.07165 [hep-ph]].



\bibitem{Dong:2015dxw} 
  P.~V.~Dong and N.~T.~K.~Ngan,
  arXiv:1512.09073 [hep-ph].



\bibitem{Cao:2015scs} 
  Q.~H.~Cao, Y.~Liu, K.~P.~Xie, B.~Yan and D.~M.~Zhang,
  Phys.\ Rev.\ D {\bf 93}, no. 7, 075030 (2016)
  doi:10.1103/PhysRevD.93.075030
  [arXiv:1512.08441 [hep-ph]].



\bibitem{Martinez:2016ztt} 
  R.~Martinez, F.~Ochoa and C.~F.~Sierra,
  arXiv:1606.03415 [hep-ph].



\bibitem{Borges:2016nne} 
  J.~S.~Borges and R.~O.~Ramos,
  Eur.\ Phys.\ J.\ C {\bf 76}, no. 6, 344 (2016)
  doi:10.1140/epjc/s10052-016-4168-8
  [arXiv:1602.08165 [hep-ph]].



\bibitem{Okada:2016whh} 
  H.~Okada, N.~Okada, Y.~Orikasa and K.~Yagyu,
  Phys.\ Rev.\ D {\bf 94}, no. 1, 015002 (2016)
  doi:10.1103/PhysRevD.94.015002
  [arXiv:1604.01948 [hep-ph]].



\bibitem{Fonseca:2016xsy} 
  R.~M.~Fonseca and M.~Hirsch,
  arXiv:1607.06328 [hep-ph].



\bibitem{Fonseca:2016tbn} 
  R.~M.~Fonseca and M.~Hirsch,
  JHEP {\bf 1608}, 003 (2016)
  doi:10.1007/JHEP08(2016)003
  [arXiv:1606.01109 [hep-ph]].



\bibitem{Pal:1994ba} 
  P.~B.~Pal,
  Phys.\ Rev.\ D {\bf 52}, 1659 (1995)
  doi:10.1103/PhysRevD.52.1659
  [hep-ph/9411406].



\bibitem{Dias:2002gg} 
  A.~G.~Dias, V.~Pleitez and M.~D.~Tonasse,
  Phys.\ Rev.\ D {\bf 67}, 095008 (2003)
  doi:10.1103/PhysRevD.67.095008
  [hep-ph/0211107].



\bibitem{Dias:2003zt} 
  A.~G.~Dias and V.~Pleitez,
  Phys.\ Rev.\ D {\bf 69}, 077702 (2004)
  doi:10.1103/PhysRevD.69.077702
  [hep-ph/0308037].



\bibitem{Dias:2003iq} 
  A.~G.~Dias, C.~A.~de S. Pires and P.~S.~Rodrigues da Silva,
  Phys.\ Rev.\ D {\bf 68}, 115009 (2003)
  doi:10.1103/PhysRevD.68.115009
  [hep-ph/0309058].



\bibitem{Mizukoshi:2010ky} 
  J.~K.~Mizukoshi, C.~A.~de S.Pires, F.~S.~Queiroz and P.~S.~Rodrigues da Silva,
  Phys.\ Rev.\ D {\bf 83}, 065024 (2011)
  doi:10.1103/PhysRevD.83.065024
  [arXiv:1010.4097 [hep-ph]].



\bibitem{Alvares:2012qv} 
  J.~D.~Ruiz-Alvarez, C.~A.~de S.Pires, F.~S.~Queiroz, D.~Restrepo and P.~S.~Rodrigues da Silva,
  Phys.\ Rev.\ D {\bf 86}, 075011 (2012)
  doi:10.1103/PhysRevD.86.075011
  [arXiv:1206.5779 [hep-ph]].



\bibitem{Cogollo:2014jia} 
  D.~Cogollo, A.~X.~Gonzalez-Morales, F.~S.~Queiroz and P.~R.~Teles,
  JCAP {\bf 1411}, no. 11, 002 (2014)
  doi:10.1088/1475-7516/2014/11/002
  [arXiv:1402.3271 [hep-ph]].



\bibitem{Cao:2015lia} 
  Q.~H.~Cao, B.~Yan and D.~M.~Zhang,
  Phys.\ Rev.\ D {\bf 92}, no. 9, 095025 (2015)
  doi:10.1103/PhysRevD.92.095025
  [arXiv:1507.00268 [hep-ph]].



\bibitem{Kubo:2004ps} 
  J.~Kubo, H.~Okada and F.~Sakamaki,
  Phys.\ Rev.\ D {\bf 70}, 036007 (2004)
  doi:10.1103/PhysRevD.70.036007
  [hep-ph/0402089].



\bibitem{Grimus:2000vj} 
  W.~Grimus and L.~Lavoura,
  JHEP {\bf 0011}, 042 (2000)
  doi:10.1088/1126-6708/2000/11/042
  [hep-ph/0008179].



\bibitem{Bora:2012tx} 
  K.~Bora,
  Horizon {\bf 2} (2013)
  [arXiv:1206.5909 [hep-ph]].



\bibitem{Xing:2007fb} 
  Z.~z.~Xing, H.~Zhang and S.~Zhou,
  Phys.\ Rev.\ D {\bf 77}, 113016 (2008)
  doi:10.1103/PhysRevD.77.113016
  [arXiv:0712.1419 [hep-ph]].



\bibitem{Alessandria:2011rc} 
  F.~Alessandria {\it et al.},
  arXiv:1109.0494 [nucl-ex].



\bibitem{Auger:2012ar} 
  M.~Auger {\it et al.} [EXO-200 Collaboration],
  Phys.\ Rev.\ Lett.\  {\bf 109}, 032505 (2012)
  doi:10.1103/PhysRevLett.109.032505
  [arXiv:1205.5608 [hep-ex]].



\bibitem{Abt:2004yk} 
  I.~Abt {\it et al.},
  hep-ex/0404039.



\bibitem{Ackermann:2012xja} 
  K.~H.~Ackermann {\it et al.} [GERDA Collaboration],
  Eur.\ Phys.\ J.\ C {\bf 73}, no. 3, 2330 (2013)
  doi:10.1140/epjc/s10052-013-2330-0
  [arXiv:1212.4067 [physics.ins-det]].



\bibitem{KamLANDZen:2012aa} 
  A.~Gando {\it et al.} [KamLAND-Zen Collaboration],
  Phys.\ Rev.\ C {\bf 85}, 045504 (2012)
  doi:10.1103/PhysRevC.85.045504
  [arXiv:1201.4664 [hep-ex]].



\bibitem{Albert:2014fya} 
  J.~B.~Albert {\it et al.} [EXO-200 Collaboration],
  Phys.\ Rev.\ D {\bf 90}, no. 9, 092004 (2014)
  doi:10.1103/PhysRevD.90.092004
  [arXiv:1409.6829 [hep-ex]].



\bibitem{Guiseppe:2011me} 
  C.~E.~Aalseth {\it et al.} [Majorana Collaboration],
  Nucl.\ Phys.\ Proc.\ Suppl.\  {\bf 217}, 44 (2011)
  doi:10.1016/j.nuclphysbps.2011.04.063
  [arXiv:1101.0119 [nucl-ex]].



\bibitem{Bilenky:2014uka} 
  S.~M.~Bilenky and C.~Giunti,
  Int.\ J.\ Mod.\ Phys.\ A {\bf 30}, no. 04n05, 1530001 (2015)
  doi:10.1142/S0217751X1530001X
  [arXiv:1411.4791 [hep-ph]].



\bibitem{Dery:2014kxa} 
  A.~Dery, A.~Efrati, Y.~Nir, Y.~Soreq and V.~Susič,
  Phys.\ Rev.\ D {\bf 90}, 115022 (2014)
  doi:10.1103/PhysRevD.90.115022
  [arXiv:1408.1371 [hep-ph]].



\bibitem{CMS:2014qxa} 
  CMS Collaboration [CMS Collaboration],
  CMS-PAS-HIG-13-034.



\bibitem{Aad:2014dya} 
  G.~Aad {\it et al.} [ATLAS Collaboration],
  JHEP {\bf 1406}, 008 (2014)
  doi:10.1007/JHEP06(2014)008
  [arXiv:1403.6293 [hep-ex]].



\bibitem{Deschamps:2009rh} 
  O.~Deschamps, S.~Descotes-Genon, S.~Monteil, V.~Niess, S.~T'Jampens and V.~Tisserand,
  Phys.\ Rev.\ D {\bf 82}, 073012 (2010)
  doi:10.1103/PhysRevD.82.073012
  [arXiv:0907.5135 [hep-ph]].



\bibitem{Bauer:2015kzy} 
  M.~Bauer, M.~Carena and K.~Gemmler,
  [arXiv:1512.03458 [hep-ph]].



\bibitem{Cabarcas:2008ys} 
  J.~M.~Cabarcas, D.~Gomez Dumm and R.~Martinez,
  Eur.\ Phys.\ J.\ C {\bf 58}, 569 (2008)
  doi:10.1140/epjc/s10052-008-0803-3
  [arXiv:0809.0821 [hep-ph]].



\bibitem{Salazar:2015gxa} 
  C.~Salazar, R.~H.~Benavides, W.~A.~Ponce and E.~Rojas,
  JHEP {\bf 1507}, 096 (2015)
  doi:10.1007/JHEP07(2015)096
  [arXiv:1503.03519 [hep-ph]].



\bibitem{Maniatis:2006fs} M.~Maniatis, A.~von Manteuffel, O.~Nachtmann and
F.~Nagel, 
Eur.\ Phys.\ J.\ C \textbf{48}, 805 (2006) [hep-ph/0605184].
\end{thebibliography}
\end{document}